\documentclass[aps,twocolumn,pra,superscript,floatfix,superscriptaddress,showpacs,longbibliography]{revtex4-1}

\usepackage{graphicx}
\usepackage{dcolumn}
\usepackage{bm}
\usepackage{color,soul}

\usepackage{amsmath}
\usepackage{amssymb}

\usepackage[colorlinks=true,%
            linkcolor=blue,%
            urlcolor=blue,%
            citecolor=blue,%
            filecolor=blue,%
            bookmarksopen=true,%
            pdfauthor={AGarrido},%
            pdftitle={JJMajorana},%
            pdfsubject={JJMbs_Manuscript},%
            pdfpagemode=UseOutlines]{hyperref}

\graphicspath{{Images/}} 

\definecolor{applegreen}{rgb}{0.55, 0.71, 0.0}
\definecolor{antiquefuchsia}{rgb}{0.57, 0.36, 0.51}
\definecolor{amethyst}{rgb}{0.6, 0.4, 0.8}

\begin{document}


\title{Majorana edge and end states in planar Josephson junctions}

\author{A. P. Garrido}
\email{a.garridohidalgo@uandresbello.edu}
 \affiliation{ 
Departamento de Ciencias Físicas, Facultad de Ciencias Exactas, Universidad Andres Bello, Sazié 2212, Santiago 837-0136, Chile.
}%
\affiliation{Department of Physics and Astronomy, Wayne State University, Detroit, Michigan 48201, USA.}

\author{P. A. Orellana}%
 \affiliation{ 
Department of Physics, Technical University Federico Santa Mar\'ia, Valparaiso, Valparaiso 2390123, Chile.
}%

\author{A. Matos-Abiague}
\affiliation{Department of Physics and Astronomy, Wayne State University, Detroit, Michigan 48201, USA.}

\date{\today}

\begin{abstract}

We theoretically investigate the localization properties of Majorana states (MSs) in proximitized, planar Josephson Junctions (JJs) oriented along different crystallographic orientations and in the presence of an in-plane magnetic field and Rashba and Dresselhaus spin-orbit couplings. We show that two types of MSs may emerge when the junction transits into the topological superconducting state. In one case, referred to as end-like MSs, the Majorana quasiparticles are mainly localized inside the normal region at the opposite ends of the junction. In contrast, edge-like MSs extend along the opposite edges of the system, perpendicular to the junction channel.
We show how the MSs can transit from end-like to edge-like and vice versa by tuning the magnetic field strength and/or the superconducting phase difference across the junction. In the case of phase-unbiased JJs the transition may occur as the ground state phase difference self-adjusts its value when the Zeeman field is varied. We propose exploiting the extended nature of edge-like MSs as effective interconnects enabling the coupling between topological states in adjacent planar JJs. The impact of electrostatic disorder on the MSs is also analyzed.

\end{abstract}

\maketitle

\section{\label{sec:intro}Introduction}

Majorana states (MSs) are zero-energy quasiparticle excitations predicted to appear localized at the boundaries of topological superconductors (TSCs) \cite{kitaev2001unpaired,kitaev2003fault,qi2011topological,leijnse2012introduction,beenakker2013search,aguado2017majorana}. The MSs obey a non-Abelian exchange statistic, which makes them promising candidates for realizing robust qubits with potential applications in fault-tolerant quantum computing \cite{ivanov2001non,nayak2008non,alicea2011non,aasen2016milestones}.

Topological superconductivity (TS) can be engineered by using semiconductor nanowires with large spin-orbit coupling (SOC) and proximitized by s-wave superconductors
\cite{PhysRevLett.105.177002,PhysRevB.82.214509,PhysRevLett.104.040502,PhysRevLett.105.077001,rokhinson2012fractional,PhysRevLett.109.227006, mourik2012signatures, das2012zero, deng2012anomalous, deng2016majorana,manna2020signature}, proximitized systems exposed to magnetic textures \cite{PhysRevLett.109.236801,PhysRevB.85.020503,PhysRevLett.117.077002,matos2017tunable,PhysRevApplied.12.034048,PhysRevB.99.134505,PhysRevB.95.140504,desjardins2019synthetic,PhysRevB.104.174502, Marra2024:JAP,Kovalev2022:JAP}, and magnetic chains on s-wave superconductors \cite{PhysRevB.84.195442,PhysRevB.85.144505,PhysRevB.88.155420,PhysRevB.88.020407,nadj2014observation,pawlak2016probing}. Proximitized planar JJs have recently emerged as promising platforms for creating and manipulating MSs \cite{PhysRevLett.89.137007,Pientka2017:PRX,PhysRevLett.111.107007,PhysRevLett.118.107701,Setiawan2019:PRBb,Setiawan2019:PRBa,PhysRevB.89.195407,fornieri2019evidence,ren2019topological,PhysRevLett.126.036802,hart2014induced,PhysRevLett.125.086802,PhysRevB.103.L180505,lesser2021phase,zhou2022fusion,PhysRevB.99.035307,PhysRevB.96.205425,PhysRevB.99.214503,PhysRevLett.124.137001,PhysRevB.102.245403,PhysRevB.101.195435,PhysRevB.104.155428,salimian2021gate,PhysRevB.95.174515,PhysRevResearch.4.043087,PhysRevB.107.245304,Pekerten2024:PRB}. In addition to the experimental advances in building such structures, proximitized planar JJs have also been shown to possess an enhanced parameter space supporting the topological superconducting state \cite{Pientka2017:PRX,Pekerten2024:APL}.

Magnetic and crystalline anisotropic effects have been predicted to appear in the Josephson junction with noncentrosymmetric materials \cite{CATSJJ,Pekerten2022:PRB,PhysRevB.56.892,PhysRevLett.105.097002}. In particular, it has been shown that in the presence of SOC the Zeeman interaction yields a strong dependence of the system properties on the magnetic field direction. Furthermore, in systems with Rashba \cite{bychkov1984oscillatory} and Dresselhaus \cite{dresselhaus1955spin} SOCs the crystallographic orientation can affect the topological superconducting state, its robustness, and signatures \cite{Scharf2019:PRB,CATSJJ,Pekerten2022:PRB}.

Most previous investigations of TS in planar JJs have focused on \emph{end} MSs, i.e., MSs that localize at the opposite ends of the junction with short localization lengths both along the junction ($\hat{y}$ direction, as shown in Fig.~\ref{fig1}) and along the system edges perpendicular to the junction ($\hat{x}$ direction, as shown in Fig.~\ref{fig1}) \cite{Pekerten2022:PRB,PhysRevLett.124.137001,CATSJJ,PhysRevB.107.035435,zhou2022fusion}. However, theoretical evidence of the existence of \emph{edge}-like MSs (i.e., MSs that are localized along the junction direction but extend along the edges perpendicular to the junction over distances significantly larger than the junction width) has been provided in previous works \cite{Sau2010:PRL,fornieri2019evidence,CATSJJ}. The formation of MSs exhibiting anomalous multilocality in three-terminal Josephson junctions has also been proposed \cite{PhysRevB.110.L041110}. Besides JJs, Majorana edge states can also emerge in hybrid superconductor/ferromagnet structures with helical magnetic textures \cite{wu2017:PRB}. Moreover, two-dimensional structures, typically associated with quantum anomalous Hall systems coupled to superconductors, have been shown to support chiral Majorana edge states \cite{Lian2018:PNAS,menard2019:Nature,Beenakker2020:SP}  flowing around the sample edges in opposite directions and Majorana corner modes localized at vertices \cite{PhysRevB.97.205134, PhysRevB.109.115413,PhysRevB.109.205158}.

In this work, we study the formation and properties of edge-like and end-like MSs in proximitized planar JJs and characterize them by introducing a quantity (here referred to as the topological gap character) that contains information about the topological charge, topological gap, and the localization nature of the zero-energy states. The norm of the topological gap character determines the size of the topological gap relative to the proximity-induced superconducting gap, and its sign indicates whether the system is in a TS state with \emph{edge}-like (positive sign) or \emph{end}-like (negative sign) MSs. We analyze how the localization character of MSs depends on relevant system parameters such as the magnetic field strength and direction, the superconducting phase difference across the junction, the SOC strength, and the junction crystallographic orientation. Moreover, our study reveals the possibility of inducing transitions from end-like to edge-like MSs (and vice versa) by tuning the magnetic field strength and/or the superconducting phase difference. In phase-unbiased JJs, the transition between end-like and edge-like MSs may occur as the ground state phase difference self-adjusts its value when the Zeeman field is varied. The paper is organized as follows. Section\ \ref{secII} presents the theoretical model and an overview of the relevant quantities used for characterizing the systems and the MSs. The numerical simulations and main results are discussed in Sec.~\ref{secIII}, while concluding remarks are given in Sec.~\ref{secIV}.

\section{Theoretical Model}\label{secII} 
\begin{figure}
    \centering\includegraphics[width=27em,height=9em]{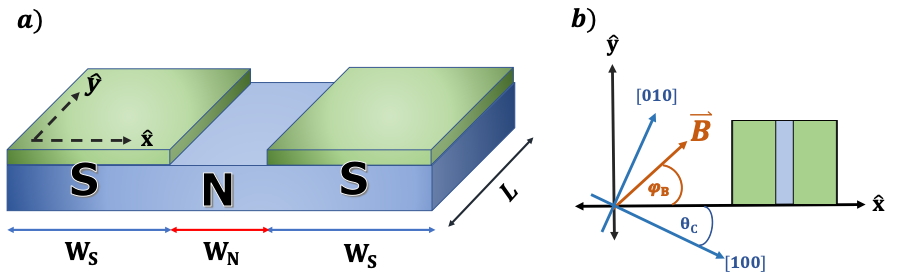}
        \caption{(a) Schematic of a JJ consisting of a noncentrosymmetric semiconductor 2DEG (blue) in contact with two superconducting (S) leads (green). The $\hat{x}$ and $\hat{y}$ axes define the coordinate system in the junction's reference frame.  A top gate (not shown) over the normal (N) region can be used to modulate the Rashba SOC strength \cite{PhysRevLett.126.036802,mayer2020gate}. (b) Relevant angles in the junction coordinate system: $\varphi_B$ defines the direction of the in-plane magnetic field ($\mathbf{B}$) with respect to the $\hat{x}$ axis, while $\theta_c$ determines the orientation of the junction reference frame with respect to the semiconductor's [100] crystallographic axis.}
    \label{fig1}
\end{figure}

We consider a planar JJ composed of a 2D electron gas (2DEG) formed in a noncentrosymmetric semiconductor and subject to an in-plane magnetic field B. The superconducting (S) regions are induced in the 2DEG by proximity to the superconducting cover layers, while the uncovered region remains in the normal (N) state [Fig.~\ref{fig1}(a)]. Excitations in the JJ are described by the Bogoliubov-de Gennes (BdG) Hamiltonian,
\begin{equation}
    H =\tau_z \otimes H_{0}+ \tau_0 \otimes \textbf{E}_z\cdot\boldsymbol{\sigma} +\Delta(x)\tau_{+}+\Delta^{*}(x)\tau_{-}\,,\label{eq1}
\end{equation}
where
\begin{widetext}
\begin{equation}
    H_{0} = \left[\dfrac{\bf{p}^2}{2m^*}+V(x)-(\mu_{S}-\varepsilon)\right]\sigma_0+\dfrac{\alpha}{\hbar}(p_y\sigma_x-p_x\sigma_y)
          +\dfrac{\beta}{\hbar}\left[(p_x\sigma_x-p_y\sigma_y)\cos 2\theta_c-(p_x\sigma_y+p_y\sigma_x)\sin 2\theta_c\right].
          \label{eq2}
\end{equation}
\end{widetext}

Here $\sigma_0$ and $\tau_0$ are unit matrices, $\sigma_{x,y,z}$ and $\tau_{x,y,z}$ denote the Pauli matrices in particle-hole and spin spaces, respectively. The linear momentum is represented by $\mathbf{p}$, $m^*$ is the electron effective mass, $\tau_{\pm}=(\tau_x\pm i\tau_y)\otimes\sigma_0/2$, and $V(x)=(\mu_S-\mu_N)\Theta(W_N/2-|x|)$ describes the difference between the chemical potentials in the N ($\mu_N$) and S ($\mu_S$) regions. The Rashba and Dresselhaus SOC strengths are represented by $\alpha$ and $\beta$, respectively. The angle $\theta_c$ characterizes the orientation of the junction with respect to the crystallographic direction [100] of the semiconductor [Fig.~\ref{fig1}-(b)]. The chemical potentials are measured with respect to the minimum of the single-particle energies, $\varepsilon=m^{*}\lambda^{2}(1+|\sin2\theta_c|)/2\hbar^{2}$. Here we use the SOC parametrization,
\begin{equation}
    \alpha = \lambda\cos\theta_{so},\ \beta=\lambda\sin\theta_{so},\ \lambda=\sqrt{\alpha^{2}+\beta^{2}} \,,\label{eq3}
\end{equation}
where $\lambda$ represents the overall strength of the combined Rashba and Dresselhaus SOCs, while the spin-orbit angle,
\begin{equation}
    \theta_{so} = \text{arccot} (\alpha/\beta) \,.\label{eq4}
\end{equation}
characterizes the relative strength between them.

The second term in Eq. (\ref{eq1}) corresponds to the Zeeman interaction and is determined by the vector,
\begin{equation}
    \textbf{E}_z =- \dfrac{g^{*}\mu_{B}B}{2}\left(\cos\varphi_B,\ \sin\varphi_B,\ 0\right)^{T} \,.\label{eq5}
\end{equation}
with $g^\ast$, $\mu_B$, $B$, and $\varphi_B$ representing the effective $g$-factor, the Bohr magneton, the magnetic field strength, and the magnetic field direction, respectively. In what follows, we use $E_Z=g^{*}\mu_{B}B/2$ to denote the amplitude of the Zeeman energy. The spatial dependence of the superconducting gap is,
\begin{equation}
    \Delta(x) =\Delta_0\, e^{i\,{\rm sgn}(x) \phi/2} \Theta(|x|-W_N/2) 
    \,,\label{eq6}
\end{equation}
where $\phi$ is the phase difference across the JJ and $\Delta_0$ is the magnitude of the proximity-induced superconducting gap.

\subsection{Topological charge}
To identify the topological regions we investigate how different sets of system parameters affect the topological invariants characterizing the junction. The presence of the magnetic field breaks the time-reversal invariance and the system generically belongs to the D class, characterized by the topological charge (i.e., the $\mathbb{Z}_2$ topological index),
\begin{equation}
    Q ={\rm sgn} \left[\dfrac{{\rm Pf}\{H(k_y=\pi)\tau_y\otimes\sigma_y\}}{{\rm Pf}\{H(k_y=0)\tau_y\otimes\sigma_y\}}\right]
    \,,\label{eq7}
\end{equation}
where ${\rm Pf}\{...\}$ denotes the Pfaffian \cite{ schnyder2008classification,ryu2010topological,ghosh2010non,Pfaffian}. The topological charge determines whether a system belonging to the D class is in a trivial ($Q=1$) or topological ($Q=-1$) phase \cite{rieder2013reentrant, adagideli2014effects, pekerten2017disorder, stanescu2011majorana, stanescu2013majorana, pekerten2019fermion}.

It is worth noting that under some conditions determined by the SOC field, the magnetic field direction, and the junction crystallographic orientation, symmetric junctions may effectively belong to the BDI class \cite{CATSJJ}. In such cases, the topological phases are characterized by the $\mathbb{Z}$ topological invariant of the BDI class. Since the topological charge $Q$ is determined by the parity of the $\mathbb{Z}$ index, the topologically non-trivial regions of the BDI class consist of non-trivial D-class regions (composed of odd $\mathbb{Z}$ index subregions) enhanced by regions with even $\mathbb{Z}$ index \cite{Pientka2017:PRX,Setiawan2019:PRBb}. We found that the topological gap in BDI class regions with multiple pairs of MSs is relatively small in the systems considered here. Therefore, our investigation focuses on regions that support only a single pair of MSs. The extent of these regions can be determined by examining how the topological charge $Q$ depends on the system parameters.

\subsection{Topological gap}

As the system transits into the topological state, MSs emerge as pairs of degenerate zero-energy states which are isolated from the rest of the excitation spectrum by the energy gap, $\Delta_{top}$, referred to as the topological gap and defined as,
\begin{equation}
    \Delta_{top}=  \left(E_1-E_0\right). \label{eq8}
\end{equation}
Here $E_0$ and $E_1$ are the two lowest-energy states on the positive branch of the energy spectrum, respectively. Due to finite-size effects, the MSs localized at opposite ends (or edges) may overlap, so their energy ($\pm E_0$) may slightly deviate from zero. Note that Eq.~(\ref{eq8}) can only be interpreted as the topological gap when the system is in the TS state.

In the topological superconducting (TS) state, the topological gap protects the Majorana bound states (MS) from smooth local perturbations. However, the degree of protection depends on the size of the topological gap, as the information stored in the MS can be compromised if the perturbation energy approaches or exceeds $\Delta_{top}$. Thus, large values of $\Delta_{top}$ are desirable for designing robust MSs suitable for constructing fault-tolerant qubits.

The magnitude of $\Delta_{top}$ has been shown to strongly depend on the junction's crystallographic orientation $(\theta_c)$, the spin-orbit angle $(\theta_{so})$ and the in-plane magnetic field orientation $(\varphi_B)$, being optimal when the following relation is fulfilled \cite{CATSJJ,Pekerten2022:PRB},
\begin{equation}\label{optimal-topogap}
    \tan\varphi_B= \cot\theta_{so}\sec 2\theta_c - \tan 2\theta_c.
\end{equation}
Therefore, it is crucial to investigate how the topological gap protecting the end-like and edge-like MSs depends on the superconducting phase difference and magnetic field strength in junctions subjected to the constraint imposed by Eq.~(\ref{optimal-topogap}).

\subsection{End-like vs. edge-like Majorana states}

As briefly discussed in the Introduction, end-like MSs are localized at the opposite ends of the junction channel, with short localization lengths both along the junction and along the system's edges perpendicular to it. In contrast, edge-like MSs are strongly localized along the junction direction but extend over distances significantly larger than the junction width along the system's edges perpendicular to the junction. We want to emphasize that the edge-like MSs examined here differ from both chiral Majorana states commonly associated with quantum anomalous Hall systems coupled to superconductors \cite{Lian2018:PNAS,menard2019:Nature,Beenakker2020:SP}  and unidirectional Majorana edge states in noncentrosymmetric superconductors \cite{Daido2017:PRB}.

To understand the origin of the extended nature of edge-like MSs, consider a simplified model of a junction sufficiently long such that the overlap between Majorana states at opposite edges is negligible. In this scenario, we can focus solely on one Majorana state, for instance, the one localized at the bottom edge ($y=0$). Assuming the S regions are infinitely wide ($W_S \to \infty$), the scattering states in the left S region can be expressed as \cite{CATSJJ,Scharf2019:PRB},
\begin{equation}\label{eq-scatt}
    \Psi(\mathbf{r})=e^{-\kappa y} \sum_{s=\pm} [C_{e,s}\chi_{e,s}e^{-iq_{e,s}x}+C_{hs}\chi_{h,s}e^{iq_{h,s}x} ],
\end{equation}
where $\kappa$ describes the localization of the Majorana state along the junction $y$-direction, the subindexes $e$ and $h$ refer to electron-like and hole-like states, respectively, $s=\pm$ characterizes the spin, and $q_{e,s}$ ($q_{h,s}$) is the wave vector of the electron-like (hole-like) state with spinor $\chi_{e,s}$ ($\chi_{h,s}$) and spin $s$. In Eq.~(\ref{eq-scatt}), we have omitted the oscillations of the MSs along the junction direction, as they become negligibly small for sufficiently long junctions.
\\
At $\phi = 0$, the coefficients $C_{e,s}$ and $C_{h,s}$ become phase-independent, and the asymptotic behavior of $\Psi(x, y)$ along the edge is entirely determined by the wave vectors. By combining Eqs.~(\ref{eq1}) and (\ref{eq-scatt}) and requiring the energy to vanish, we obtain the following quartic equation for $q^2$,
\begin{widetext}
\begin{equation}\label{eq-quartic}
    \left(|\mathbf{E}_Z|^2+\mathbf{w}\cdot\mathbf{w}+|\Delta_0|^2+\xi^2\right)^2-4\left[|\mathbf{E}_Z|^2(|\Delta_0|^2+\xi^2)+(\mathbf{w}\cdot\mathbf{w})\xi^2+(\mathbf{E}_z\cdot\mathbf{w})^2\right]=0,
\end{equation}
\end{widetext}
where,
\begin{equation}\label{eq-sof}
\mathbf{w}=
\begin{pmatrix}
q\beta\cos2\theta_c+i\kappa(\alpha-\beta\sin2\theta_c) \\
-q(\alpha+\beta\sin2\theta_c)-i\kappa\beta\cos2\theta_c \\
0
\end{pmatrix},
\end{equation}
is the spin-orbit field and $\xi=\hbar^2q^2/(2m^\ast)-\mu_S+\epsilon$.
\\
The solutions to the quartic equation [Eq.~(\ref{eq-quartic})] represent the wave vectors \(q_{j,s}\) (\(j=e,h\)). At \(\phi=0\), the Majorana state’s character is determined by the nature of these wave vectors. Specifically, each scattering mode $\{j, s\}$ contributing to the Majorana state in the S region decays within a characteristic length,
\begin{equation}
    l_{j,s}\sim\frac{1}{{\rm Im}[q_{j,s}]}\;.
\end{equation}
The overall localization length of the Majorana state along the edge is given by the largest of the $l_{j,s}$, which may become significantly larger than the junction width for wave vectors with a sufficiently small imaginary part, leading to the formation of edge-like MSs. At $\phi=0$, if at least one wave vector is purely real, the corresponding scattering mode represents a propagating wave with infinite localization length. In this particular case, the edge-like MS propagates without decay over the entire edge. Conversely, end-like MSs appear when all wave vectors are complex.
\\
The coefficients of the quartic equation [Eq.~(\ref{eq-quartic})] are generally complex, yielding complex solutions for all wave vectors and resulting in end-like MSs. However, if the conditions,
\begin{equation}\label{eq-cond-real}
 \alpha^2\sin2\varphi_B+\beta^2\sin(2\varphi_B+4\theta_c)=\alpha\beta\cos2\theta_c=0,
\end{equation}
are satisfied, all coefficients of the quartic equation become real. This allows for purely real wave vectors, enabling the emergence of propagating edge-like MSs if the system parameters are appropriately tuned. Note that Eq.~(\ref{eq-cond-real}) is a necessary, though not sufficient, condition for the emergence of propagating edge-like MSs in junctions with infinitely wide S regions.

In general, we can anticipate three scenarios: i) four real wave vectors, leading to edge-like MSs composed of the superposition of four propagating modes; ii) two real and two complex wave vectors, resulting in edge-like MSs formed by two propagating and two localized modes; and iii) four complex wave vectors, corresponding to localized end-like MSs. By appropriately tuning the system parameters, the system can be driven into any of these scenarios, enabling controlled transitions between end-like and edge-like behaviors. Furthermore, at finite superconducting phase differences, the coefficients $C_{e,s}$ and $C_{h,s}$ in Eq.~(\ref{eq-scatt}) become $\phi$-dependent.
This dependence governs the interference between the four scattering modes, thereby modulating the localization characteristics of the Majorana states (MSs).

Propagating edge-like MSs with infinite localization length cannot be practically realized, as their extension is ultimately limited by the finite size of the S region in realistic JJs. Therefore, we focus our numerical investigation on edge-like MSs with negligibly small decay over the length of the S regions, which, in the numerical simulations, are set to be much larger than the junction width. To ensure that the edge-like MSs extend across the entire S regions of the finite-size junctions considered, we impose the necessary constraints derived from the asymptotic analysis in the limit $W_S\rightarrow\infty$. Consequently, all configurations in the numerical analysis presented in Sec.~\ref{secIII} are chosen such that Eq.~(\ref{eq-cond-real}) and the condition for optimal topological gap [Eq.~(\ref{optimal-topogap}) are simultaneously satisfied.

To capture the topological character of the state, the magnitude of the topological gap, and the extension of the MSs along the edges perpendicular to the junction, we introduce the quantity,
\begin{equation}
    \tilde{\Delta} = \dfrac{(1-Q)\Delta_{top}}{2\Delta_0}\,{\rm sgn}(\eta)\label{eq10}
\end{equation}
where $\eta$ is a parameter that characterizes the nature of the MSs, taking positive values for edge-like MSs and negative values for end-like MSs.

In what follows, we refer to $\tilde{\Delta}$ as the \emph{topological gap character}. Its magnitude represents the topological gap normalized to the proximity-induced superconducting gap, while its sign indicates the nature of the Majorana states: positive for edge-like MSs, negative for end-like MSs, and zero when the system is in the trivial state (or in the BDI-class TS with an even number of Majorana pairs), i.e.,
\begin{equation}
    \tilde{\Delta} = 
    \begin{cases}
            +\Delta_{top}/\Delta_0 & \text{for D-class TS with edge-like MSs} \\
      0 & \text{for trivial phase with no MSs} \\
      0 & \text{for even BDI-class TS phase}\\
      -\Delta_{top}/\Delta_0 & \text{for D-Class with end-like MSs} 
    \end{cases}.\label{eq11}
\end{equation}
Although we limit our analysis to MSs within the D class, the topological character defined in Eq.~(\ref{eq10}) can easily be generalized to account for the BDI-class TS by using the winding number (the $\mathbb{Z}$ topological invariant) instead of the topological charge.
\\
For the junctions studied here, the probability density of edge-like MSs displays an oscillatory behavior along the edge, with multiple local maxima of comparable amplitudes. In contrast, the probability density oscillations of end-like MSs decay rapidly, featuring a large maximum at the middle of the junction edge, followed by a few significantly smaller maxima. Therefore, the localization nature of the MSs can be captured by the parametrization $\eta =n-n_0-1/2$, where $n$ is the number of probability density local maxima with comparable amplitudes. Since the typical number of local maxima for end-like MSs is smaller than 3, we set $n_0=3$ in the numerical simulations of Eq.(\ref{eq11}) discussed below. For higher chemical potentials and/or larger $W_N$ and $W_S$ one may need to increase $n_0$ accordingly. Note that other equivalent definitions of $\eta$ are also possible. For instance, one could define $\eta$ to be positive when the MS localization length is much larger than the edge length $W$, and negative otherwise. However, the two approaches are equivalent, since local maxima of the probability density extend along the entire edge with comparable amplitudes only when the localization length significantly exceeds $W$.

\subsection{Phase-biased and phase-unbiased JJs}

The eigenenergies $E_n$ can be used to compute the phase-dependent part of the junction's free energy,
\begin{equation}
    F=-\sum_{E_n > 0}[E_n+2k_{B}T\ln{(1+e^{-E_n/k_B T})}].\label{eq12}
\end{equation}
In the phase-biased case, the JJ is incorporated in a closed loop threaded by a magnetic flux $\Phi$, which fixes the superconducting phase across the junction to $\phi=2\pi\Phi/\Phi_{0}$, where $\Phi_{0}$ is the magnetic flux quantum.

In the absence of the magnetic flux, the junction is phase unbiased, and the phase difference is self-adjusted in such a way that the free energy of the system is minimized. The ground-state phase ($\phi_{GS}$) is the superconducting phase difference that minimizes the free energy of the system, i.e.,
\begin{equation}
    F(\phi_{GS})= \underset{\phi}{\text{min}} \hspace{0.5em} F(\phi).\label{eq13}
\end{equation}
Since the free energy also depends on the Zeeman energy, $\phi_{GS}$ is generally a function of the magnetic field. This offers a mechanism for indirectly controlling the superconducting phase difference using an in-plane magnetic field, without relying on a magnetic flux.

\section{Results}\label{secIII}

We consider two types of junctions: (i) Al/HgTe Josephson junctions (JJs), where Rashba spin-orbit coupling (SOC) is the dominant effect, and (ii) Al/InSb JJs, where both Rashba and Dresselhaus SOC may be significant. The system parameters used in our calculations are provided in Appendix \ref{AppendixA}. The numerical simulations were carried out by discretizing Eq.~(\ref{eq1}) on a mesh with a lattice constant of $a=10$ nm. Using the finite-difference approximation, we constructed the tight-binding (TB) form of the BdG Hamiltonian with the Kwant package \cite{groth2014kwant}. The energy spectrum and wave functions were obtained by numerically diagonalizing the TB BdG Hamiltonian on a finite lattice. The energy spectrum was then used to compute the topological gap [Eq.~(\ref{eq8})], the phase-dependent part of the free energy [Eq.~(\ref{eq11})], and the ground-state phase [Eq.~(\ref{eq11})]. The topological charge [see Eq.~(\ref{eq7})] was computed by using the system TB BdG Hamiltonian with imposed translational invariance along the junction direction.

\begin{figure}
    \centering
    \includegraphics[width=27em,height=23em]{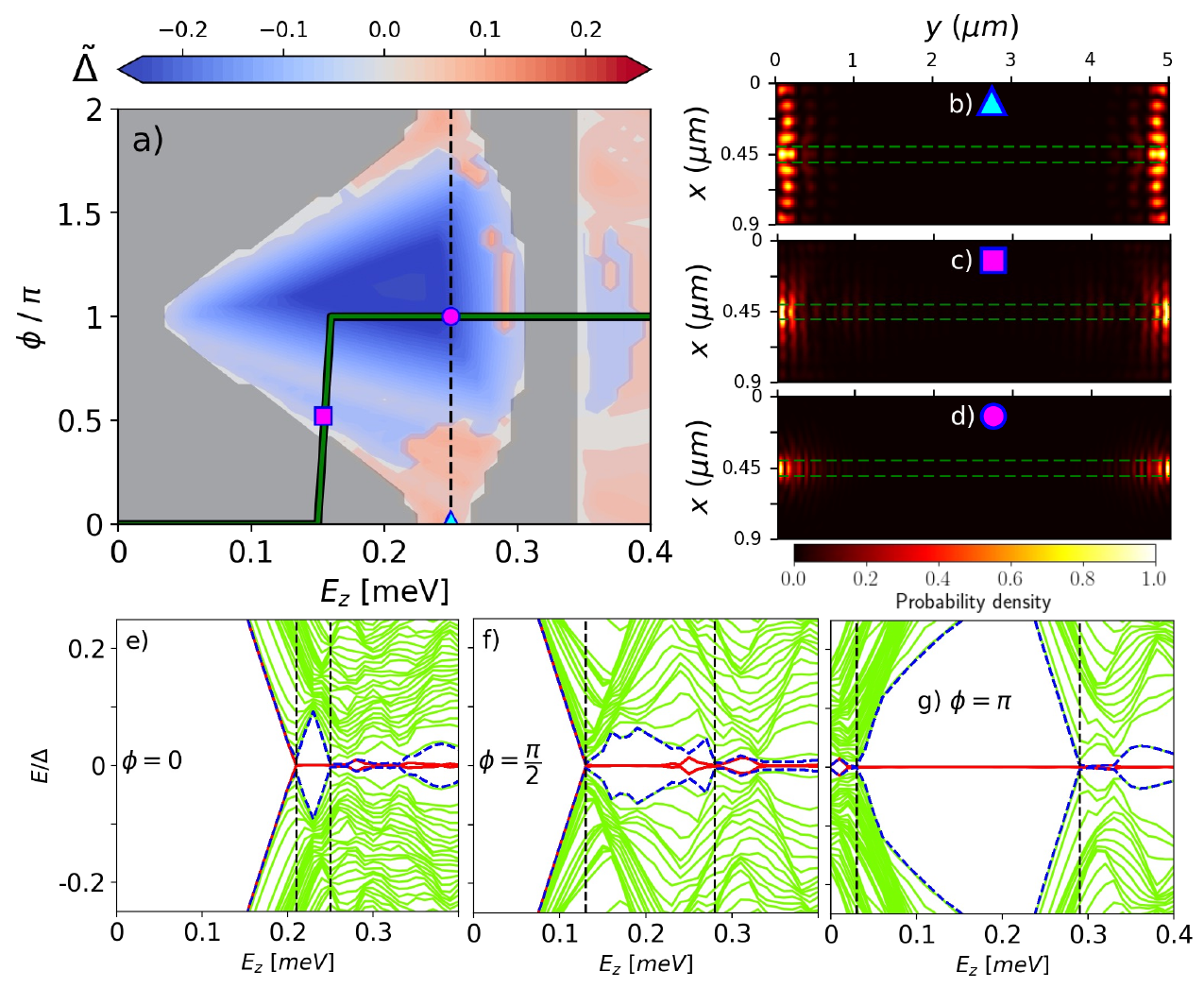}
    \caption{(a) Topological gap character ($\tilde{\Delta}$) as a function of the Zeeman energy $E_Z$ and the superconducting phase difference ($\phi$) across an Al/HgTe JJ with only Rashba SOC ($\theta_{so}=0$). The junction and magnetic field orientations are set to $\theta_c=0$ and $\varphi_B=\pi/2$, respectively. The green solid line represents the path of the ground-state phase ($\phi_{GS}$) as the Zeeman energy is varied. The vertical dashed line marks a possible transition between a TS state supporting a zero-phase edge-like MS (cyan triangle) and one supporting an end-like MS (magenta dot) during which $E_Z$ is kept constant, while $\phi$ is tuned. (b)-(d) Probability density (normalized to its maximum value) of the MSs corresponding to the $E_Z$ and $\phi$ values marked in (a) by the cyan triangle (edge-like MS), magenta square (end-like MS), and magenta dot (end-like MS), respectively. (e)-(g) Energy spectra as a function of the Zeeman energy for $\phi=0$, $\phi=\pi/2$, and $\phi=\pi$, respectively. Red-solid and dashed-blue lines represent states that evolve into MSs as $E_Z$ is varied. Vertical dashed lines indicate the boundaries of the first topological region in which only a single pair of MSs (red solid lines) exists.} \label{fig2}
\end{figure}

\subsection{Effects of Rashba SOC}

In Al/HgTe Josephson junctions (JJs), where Rashba SOC is the dominant and Dresselhaus SOC is negligibly small (i.e., $\beta\approx 0$), the spin-orbit angle $\theta_{so}\approx 0$ and $\lambda\approx\alpha$ [see Eqs.~(\ref{eq3}) and (\ref{eq4})]. In this scenario, the system exhibits magneto anisotropyy (i.e., its properties depend on the in-plane magnetic field orientation, $\varphi_B$), while crystalline anisotropy is absent (i.e., the system properties are independent of the junction's crystallographic direction, $\theta_c$) \cite{CATSJJ,Pekerten2022:PRB,Pekerten2024:APL}. For the calculations presented in this subsection, the in-plane magnetic field was aligned with the junction direction (i.e., $\varphi_B=\pi/2$), which, as discussed in Ref.~\cite{CATSJJ}, yields the optimal topological gap for this configuration.

Figure \ref{fig2}(a) shows the behavior of the topological gap character ($\tilde{\Delta}$), as a function of the Zeeman field ($E_Z$) and the superconducting phase difference ($\phi$) in a phase-biased JJ. Gray areas correspond to $\tilde{\Delta}=0$, indicating topologically trivial (with respect to the D classification) regions with topological charge $Q=1$. Note, however, that the gray zone may still contain regions of BDI-class TS with an even number of MS pairs. As previously mentioned here we disregard those regions, as they exhibit a relatively small topological gap. Blue (red) areas indicate regions where $\tilde{\Delta}<0$ ($\tilde{\Delta}>0$) correspond to a D-class TS phase that supports the formation of end-like (edge-like) MSs. The figure reveals that in junctions where only Rashba SOC is significant, the formation of end-like MSs with a sizable topological gap is favored for $\phi$-values near $\pi$. In contrast, less robust edge-like MSs with smaller topological gaps emerge when the system is in the TS state, and $\phi$ is near 0 or $2\pi$.

The effect of the superconducting phase difference on the localization nature of the MSs can be qualitatively understood by noting that the phase factor of the superconducting pairing potential appearing in the antidiagonal blocks of the BdG Hamiltonian can be gauged away by a position-dependent unitary transformation (see details in Appendix \ref{AppendixB}). As a result, the BdG Hamiltonian of a junction with a superconducting phase difference $\phi$ is transformed into a BdG Hamiltonian of a junction with zero phase difference, but in the presence of a position-dependent gauge potential with strength proportional to $\phi$. The edge-like MSs at zero phase transit to end-like MSs when the $\phi$-dependent gauge field is strong enough. 

Considering a junction with infinitely wide S regions (i.e., $W_S\rightarrow \infty$) and a Zeeman field along the junction direction and much larger than the Rashba SOC splitting, we can write the approximate solutions of Eq.~(\ref{eq-quartic}) as,
\begin{equation}\label{eq-qvec}
q_{j,s}=\sqrt{\left[\dfrac{\sqrt{2m^{*}\left(\mu_S+j\sqrt{E_Z^2 -\Delta_0^2}\right)}}{\hbar} +s \;k_{so}\right]^2+\kappa^2}
\end{equation}
where $j=e=1$ ($j=h=-1$) for electron-like (hole-like) states, $E_Z=|\mathbf{E}_Z|$, and $k_{so}=m^\ast\alpha/\hbar^2$. This indicates that zero-phase propagating edge-like MSs of junctions with only Rashba SOC and magnetic field along the junction direction appear when the system is in the TS state and the following conditions are fulfilled,
\begin{equation}\label{edge-condition}
    E_Z\geq \Delta_0\;{\rm and}\;\mu_S\geq \sqrt{E_Z^2-\Delta_0^2}.
\end{equation}
Indeed, in such a situation, all the wave vectors are purely real (i.e. ${\rm Im}[q_{j,s}]=0$) and the the MS localization length, $l_{j,s}\to\infty$.

When the second inequality in Eq.(\ref{edge-condition}) is satisfied (as it is for the parameters used in the numerical simulations), the emergence of edge-like MSs at $\phi=0$ in the limit $W_S\rightarrow \infty$ is governed by the condition $E_Z\geq \Delta_0$. However, since the numerical simulations account for superconducting regions of finite width, the results shown in Fig.~\ref{fig2}(a) —where zero-phase edge-like MSs appear at Zeeman energies slightly below $\Delta_0$ —exhibit a small deviation from the condition $E_Z\geq \Delta_0$. The impact of finite-size effects on the topological phase diagram of planar JJs has been explored in Ref.\cite{Pekerten2024:APL}.

In phase-unbiased junctions, the system's state evolves according to the trajectory of the ground-state phase [see green solid line in Fig.~\ref{fig2}(a)] as the Zeeman field is varied. During the $0-\pi$ ground-state jump at $E_Z\approx0.14$ meV, the junction undergoes a transition from the trivial to a TS phase with end-like MSs.
A self-tuning mechanism, where edge-like MSs transition into end-like MSs (and/or vice versa) as the Zeeman field varies, appears impractical in phase-unbiased junctions with only Rashba SOC.
However, a topologically protected transition in which end- and edge-like MSs can be transformed into each other without exiting the TS state seems feasible by tuning the magnetic flux in phase-biased Josephson junctions (JJs). Since the system remains in the same TS state the topological gap does not close during the process, providing topological protection to the transition. This is illustrated in Fig.~\ref{fig2}(a), where an edge-like MS (cyan triangle) can evolve into an end-like MS (magenta dot) by adjusting the Zeeman field (at the value indicated by the vertical dashed line) and varying the phase from $0$ to $\pi$.

To visualize the spatial extension of the edge-like MSs along the sample edges perpendicular to the junction, Fig.~\ref{fig2}(b) shows the probability density (normalized to its maximum value) of the edge-like MS corresponding to the cyan triangle in Fig.~\ref{fig2}(a) at $\phi=0$. The probability density exhibits an oscillatory behavior with multiple maxima of comparable amplitude. Note that the edge-like MSs consistently spread along the entire edge extension (see Appendix \ref{AppendixC}). In contrast, the probability density of the end-like MSs corresponding to the magenta square and magenta dot in Fig.~\ref{fig2}(a) [see Figs.~\ref{fig2}(c) and (d), respectively] is localized at the opposite ends of the junction with wave functions that decay both along and perpendicular to the junction's direction.

The energy spectrum as a function of $E_Z$ is shown in Figs.~\ref{fig2}(e)-(g) for $\phi=0,\pi/2,\pi$, respectively. The red solid and blue dotted lines represent the states with energies closest to zero, which evolve into MS as the junction transitions into the TS phase. The vertical lines mark the Zeeman energy boundaries of the first topological region, characterized by the $\mathbb{Z}_2$ invariant of the D class, where the topological charge is $Q=-1$ [see Eq.~(\ref{eq7})]. Within this region, the absolute value of the $\mathbb{Z}$ invariant of the BDI class equals 1, indicating the presence of a single pair of MSs, as shown by the red lines in Figs.~\ref{fig2}(e)-(g). Additionally, regions where a second pair of MSs emerge (blue dashed lines) can also be observed. In these regions, the system remains in the BDI class and supports two pairs of MSs, corresponding to an absolute value of 2 for the $\mathbb{Z}$ invariant of the BDI class. However, it is important to note that the topological gap in regions with multiple pairs of MSs is significantly smaller than in regions where only a single pair of MSs exists [this is particularly clear in Fig.~\ref{fig2}(g)]. For this reason, our analysis of edge-like and end-like MSs is focused on topological regions containing only a single pair of MSs.


\begin{figure}
    \centering
    \includegraphics[width=27em,height=23em]{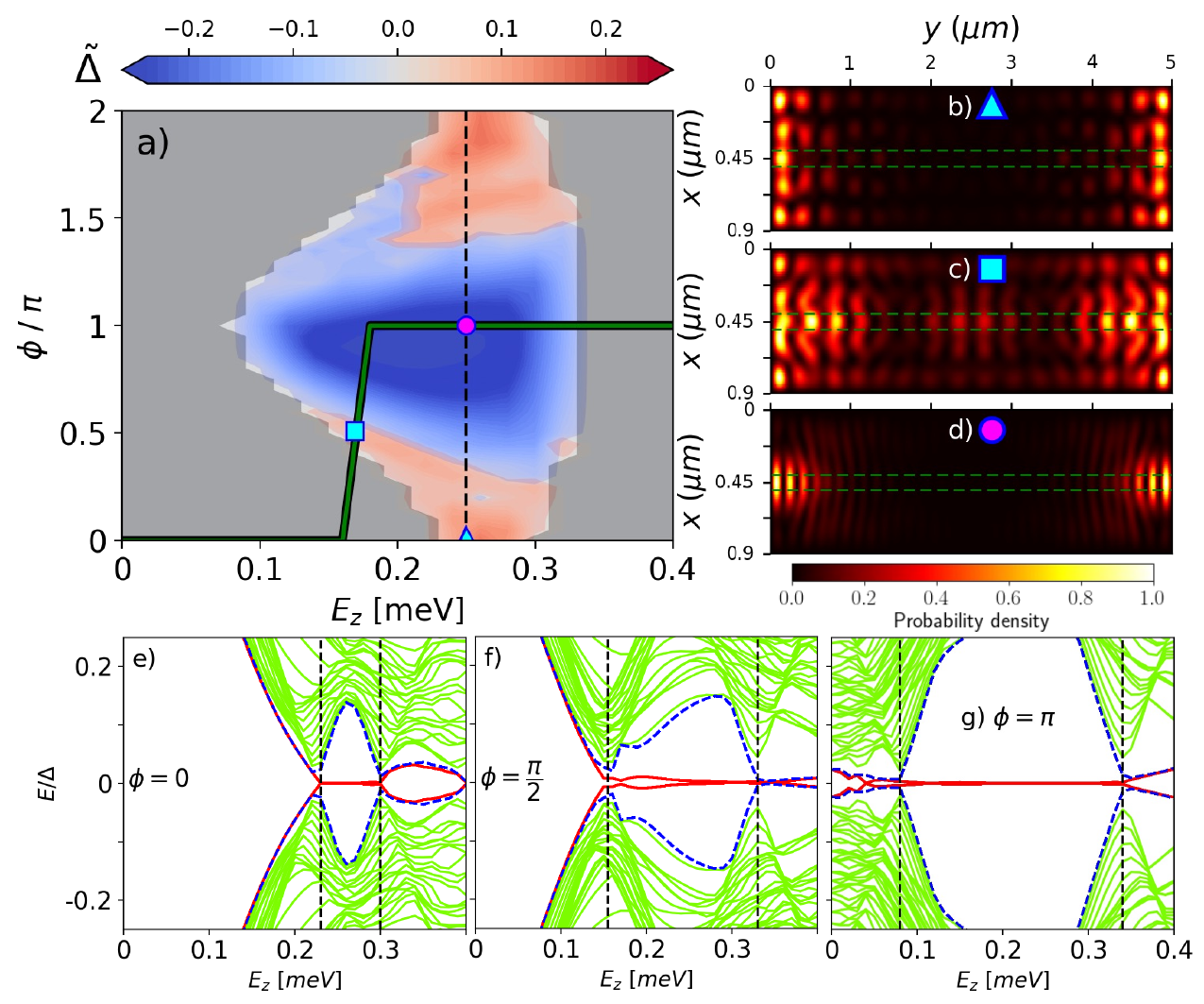}
    \caption{(a) Topological gap character ($\tilde{\Delta}$) as a function of the Zeeman energy $E_Z$ and the superconducting phase difference ($\phi$) across an Al/InSb JJ, where the Rashba SOC has been tuned to a negligibly small value and only Dresselhaus SOC is relevant ($\theta_{so}=\pi/2$). The junction and magnetic field orientations are set to $\theta_c=0=\varphi_B=0$. The green solid line represents the path of the ground-state phase ($\phi_{GS}$) as the Zeeman energy is varied. The vertical dashed line marks a possible transition between a TS state supporting a zero-phase edge-like MS (cyan triangle) and one supporting an end-like Majorana state (magenta dot) during which $E_Z$ is kept constant, while $\phi$ is tuned. (b)-(d) Probability density (normalized to its maximum value) of the MSs corresponding to the $E_Z$ and $\phi$ values marked in (a) by the cyan triangle (edge-like MS), cyan square (edge-like MS), and magenta dot (end-like MS), respectively. (e)-(g) Energy spectra as a function of the Zeeman energy for $\phi=0$, $\phi=\pi/2$, and $\phi=\pi$, respectively. Red-solid and dashed-blue lines represent states that evolve into MSs as $E_Z$ is varied. Vertical dashed lines indicate the boundaries of the first topological region in which only a single pair of MSs (red solid lines) exists. The deviation of the edge-like MS energies (red solid lines) from zero in (f) results from the wavefunction overlap between edge-like MSs on opposite edges [see (c)].}
    \label{fig3}
\end{figure}

\subsection{Effects of Dresselhaus SOC}

We now focus on JJs where only the Dresselhaus SOC plays a significant role. In this scenario, with $\alpha=0$ and $\lambda =\beta\neq 0$, we can assume $\theta_{so}=\pi/2$ without loss of generality. Unlike linear Rashba SOC, the Dresselhaus spin-orbit field is not rotationally invariant about the axis normal to the junction plane, leading to the emergence of both magneto-anisotropy and crystalline anisotropy in the system. Therefore, to optimize the topological gap, it is essential to carefully align both the in-plane magnetic field angle $\varphi_B$ and the crystallographic orientation of the junction $\theta_c$. For numerical simulations, we use the values $\theta_c =\varphi_B = 0$, which together with $\theta_{so}=\pi/2$, satisfy the condition in Eq.~(\ref{optimal-topogap}).


The character of the topological gap $\tilde{\Delta}$, depicted in Fig.~\ref{fig3}(a) as a function of $E_Z$ and $\phi$, exhibits an overall behavior similar to that shown in Fig.~\ref{fig2}(a). Specifically, edge and end-like MSs emerge at phase values near 0 and $\pi$, respectively. Notably, edge-like MSs are better protected by a larger topological gap in JJs dominated by Dresselhaus SOC compared to those where Rashba SOC is predominant.

The ground-state phase trajectory [green line in Fig.~\ref{fig2}(a)] in phase-unbiased JJs undergoes a $0-\pi$ transition, enabling the system to enter the TS state at a lower Zeeman energy compared to when the phase is fixed at zero. However, as in the case of Rashba SOC, a self-tuned transition from end-like to edge-like MSs appears unfeasible in JJs with only Dresselhaus SOC. Nonetheless, the end-to-edge transition can still be achieved with external control of the superconducting phase difference. For example, as indicated by the vertical dotted line in Fig. \ref{fig3}(a), fixing the in-plane magnetic field to a value corresponding to $E_Z\approx 0.25$~meV and tuning $\phi$ from 0 to $\pi$ would induce a transition from edge to end-like MSs.

The probability density of MSs corresponding to the values of $E_Z$ and $\phi$ marked by the cyan triangle, cyan square, and magenta dot in Fig.~\ref{fig3}(a) are shown in Figs.~\ref{fig3}(b)-(d), respectively. Additionally, the dependence of the energy spectra on $E_Z$ for $\phi=0,\pi/2,\pi$ is depicted in Figs.~\ref{fig3}(e)-(g), respectively. Compared to the edge-like MS illustrated in Fig.~\ref{fig2}(b), the zero-phase edge-like MSs in Fig.~\ref{fig3}(b) are protected by a larger topological gap [see Figs.~\ref{fig2}(e) and \ref{fig3}(e)]. However, MSs in junctions with only Rashba SOC exhibit stronger localization along the junction, implying that JJs with dominant Dresselhaus SOC would need to be longer to effectively prevent the overlap between MSs localized at opposite ends (or edges). The overlap between MSs from opposite edges, particularly evident in Fig.~\ref{fig3}(c), causes their energies to deviate from zero [seen in Fig.~\ref{fig3}(f)].


\subsection{Effects of combined Rashba and Dresselhaus SOCs }

In systems like Al/InSb-based junctions, the coexistence of significant Rashba and Dresselhaus SOC generates a two-fold symmetric spin-orbit field. This, combined with the Zeeman interaction, results in nontrivial magneto anisotropic and crystalline anisotropic effects. The relative strength of Rashba ($\alpha$) and Dresselhaus ($\beta$) SOCs can be tuned by adjusting the Rashba SOC via a gate placed on top of the junction \cite{PhysRevLett.126.036802}.
A particularly interesting regime arises when the Rashba and Dresselhaus SOCs are equally strong. Based on the parametrization introduced in Eqs.~(\ref{eq3}) and (\ref{eq4}), the condition $\alpha=\beta$ corresponds to the SOC angle $\theta_{so}=\pi/4$.

\begin{figure}[t]
    \centering
    \includegraphics[width=27em,height=23em]{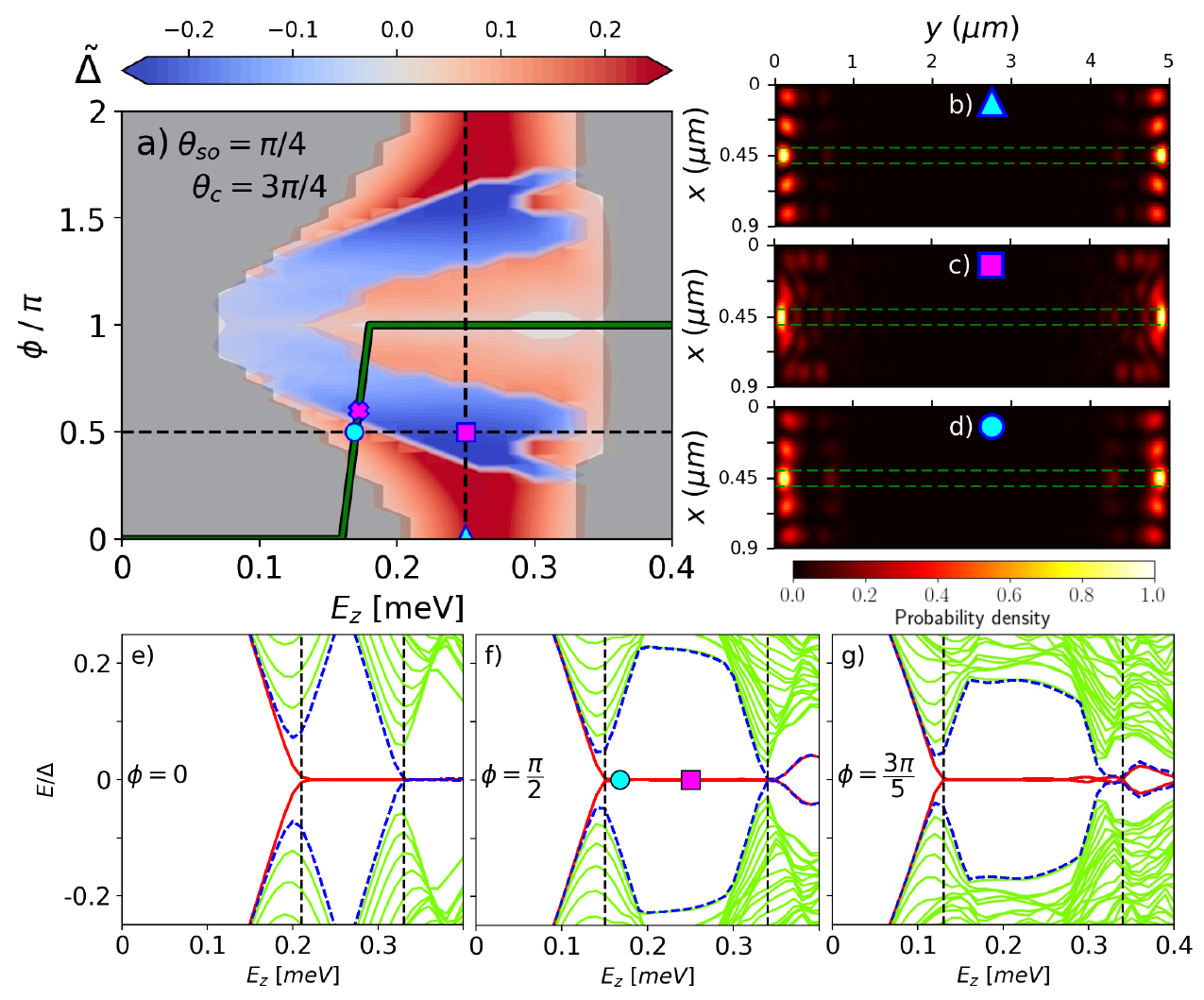}
     \caption{(a) Topological gap character ($\tilde{\Delta}$) as a function of the Zeeman energy $E_Z$ and the superconducting phase difference ($\phi$) across an Al/InSb JJ, with equal Rashba and Dresselhaus SOC strengths ($\theta_{so}=\pi/4$). The junction and magnetic field orientations are set to $\theta_c=3\pi/4$ and $\varphi_B=\pi/2$, respectively. The green solid line represents the path of the ground-state phase ($\phi_{GS}$) as the Zeeman energy is varied. The vertical (horizontal) dashed line marks a possible transition between a TS state supporting a zero-phase edge-like MS (cyan triangle/dot) and one supporting an end Majorana state (magenta square) during which $E_Z$ ($\phi$) is kept constant while $\phi$ ($E_Z$) is tuned. A transition between edge-like (e.g., cyan dot) and end-like (e.g., magenta cross) MSs can also be achieved by solely tuning $E_Z$, as the value of $\phi$ self-adjusts and follows the path of the ground-state phase (green solid line). (b)-(d) Probability density (normalized to its maximum value) of the MSs corresponding to the $E_Z$ and $\phi$ values marked in (a) by the cyan triangle (edge-like MS), magenta square (end-like MS), and cyan dot (edge-like MS), respectively. (e)-(g) Energy spectra as a function of the Zeeman energy for $\phi=0$, $\phi=\pi/2$, and $\phi=3\pi/5$, respectively. Red-solid and dashed-blue lines represent states that evolve into MSs as $E_Z$ is varied. Vertical dashed lines indicate the boundaries of the first topological region in which only a single pair of MSs (red solid lines) exists.}
    \label{fig4}
\end{figure}

The character of the topological gap ($\tilde{\Delta}$) is shown in Fig.~\ref{fig4}(a) as a function of $E_Z$ and $\phi$, with $\theta_{so}=\pi/4$. The crystallographic and magnetic field orientations were chosen as $\theta_c=3\pi/4$ and $\varphi_B=\pi/2$, respectively, ensuring that the condition in Eq.~(\ref{optimal-topogap}) is satisfied. A distinctive feature of this regime is that, unlike the previously discussed cases, edge-like MSs can emerge at superconducting phase differences close to $\pi$. Additionally, the coexistence of Rashba and Dresselhaus SOCs leads to an overall enhancement of the topological gap.

As in the cases discussed in the previous subsections, in phase-biased JJs with both Rashba and Dresselhaus SOC transitions between edge and end-like MSs can also be induced by tuning the magnetic field while keeping the phase fixed at an appropriate value [e.g., following the dotted line from the cyan dot to the magenta square in Fig.~\ref{fig4}(a)] or by fixing the magnetic field and tuning the phase difference [e.g, following the dotted line from the cyan triangle to the magenta square in Fig.~\ref{fig4}(a)]. Remarkably, the coexistence of Rashba and Dresselhaus SOC allows for transitions between edge and end-like MSs in phase-unbiased junctions, something not observed in systems with only Rashba or only Dresselhaus SOC. As shown in Fig.~\ref{fig4}(a), in a phase-unbiased JJ the self-tuning of the ground-state phase (green solid line) as the Zeeman energy varies, creates a transition pathway between (red) regions hosting edge-like MSs and (blue) regions hosting end-like MSs. For instance, along the ground-state trajectory, the edge-like MSs at the cyan dot can transition into the end-like MSs at the magenta cross. Further increasing the Zeeman field induces a $0$–$\pi$ jump in the ground-state phase, causing edge-like MSs to re-emerge.

For completeness, the probability densities of the MSs corresponding to the $E_Z$ and $\phi$ values indicated by the cyan triangle, magenta square, and cyan dot in Fig.~\ref{fig4} are shown in Figs.~\ref{fig4}(b)-(d), respectively. Both the edge-like MSs [Figs.~\ref{fig4}(b) and (d)] and the end-like MSs [Fig.~\ref{fig4}(c)] exhibit strong localization along the junction direction, resulting in very stable MSs within the first topological region, which contains a single pair of MSs. This stability is evident in the energy spectra presented in Figs.~\ref{fig4}(e)-(g) for $\phi=0,\pi/2,3\pi/5$, where very flat zero-energy MSs (red lines) with an enhanced topological gap, compared to the cases with only Rashba SOC [see Figs.~\ref{fig2}(e)-(g)] or only Dresselhaus SOC [see Figs.~\ref{fig3}(e)-(g)], are clearly visible. Notably, the transition between edge-like (cyan dot) and end-like (magenta square) MSs induced by varying $E_Z$ while keeping $\phi=\pi/2$ is very robust, with MSs maintaining their zero energy and protected by a sizable topological gap, as can be appreciated in Figs.~\ref{fig4}(f).


\begin{figure}[t]
    \centering
    \includegraphics[width=27em,height=23em]{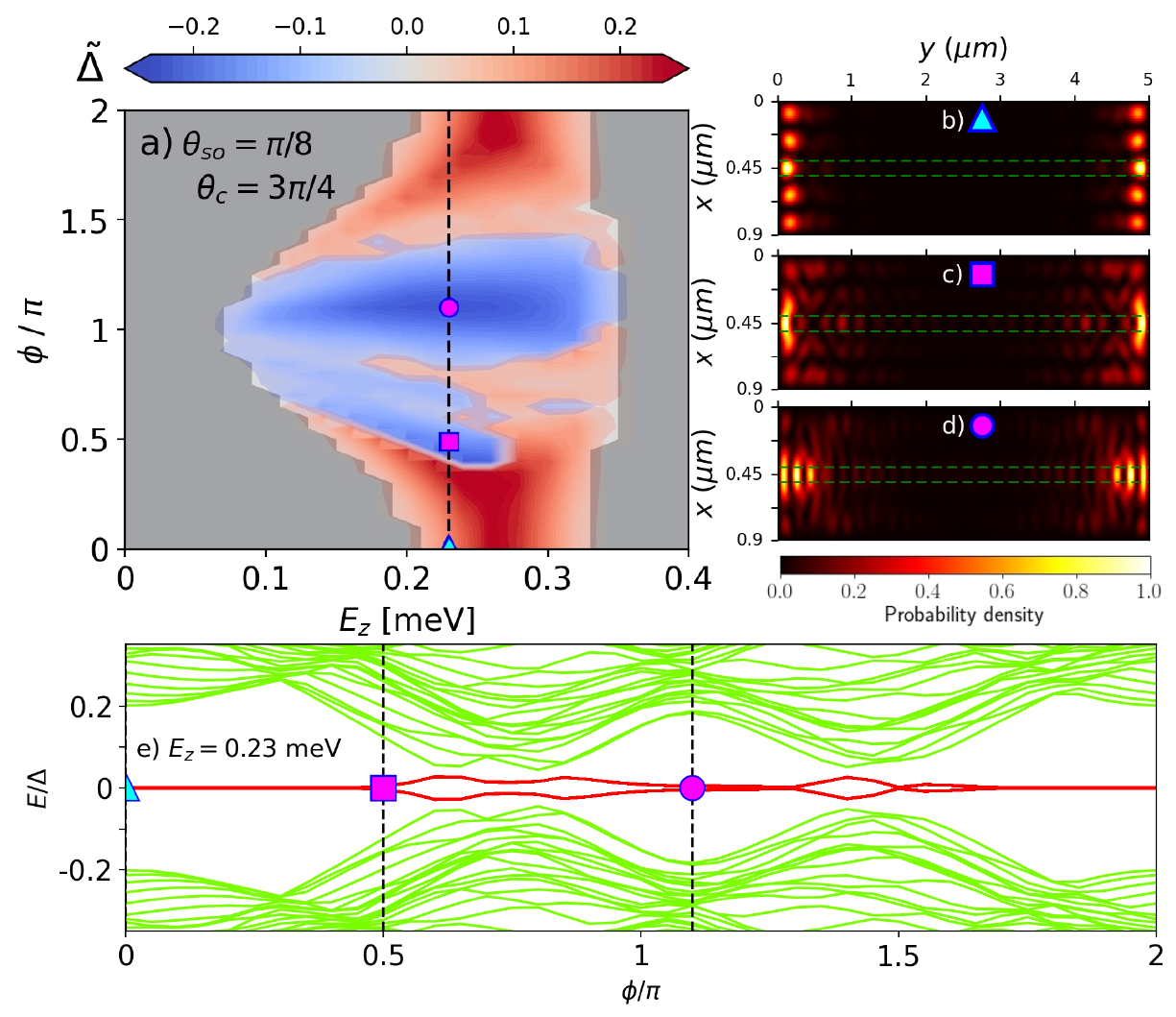}
    \caption{(a) Topological gap character ($\tilde{\Delta}$) as a function of the Zeeman energy $E_Z$ and the superconducting phase difference ($\phi$) across an Al/InSb JJ, with Rashba SOC strength about 2.4 times greater than the Dresselhaus SOC strength ($\theta_{so}=\pi/8$). The junction and magnetic field orientations are set to $\theta_c=3\pi/4$ and $\varphi_B=\pi/2$, respectively. (b)-(d) Probability density (normalized to its maximum value) of the MSs corresponding to the $E_Z$ and $\phi$ values marked in (a) by the cyan triangle (edge-like MS), magenta square (end-like MS), and magenta dot (end-like MS), respectively. (e) Energy spectrum along the path indicated by the vertical dashed line in (a), where $E_Z=0.23$~meV and $\phi$ is varied from 0 to $2\pi$. The symbols in (e) indicated the energy of the MSs whose probability densities are plotted in (b)-(d).}
    \label{fig5}
\end{figure}

In junctions where the strengths of Rashba and Dresselhaus SOC are equal (e.g., $\theta_{so}=\pi/4$), the topological gap exhibits mirror symmetry with respect to $\phi = \pi$, as shown in Fig.~\ref{fig4}(a). However, this symmetry is broken when the Rashba and Dresselhaus SOC strengths are no longer equal,  as illustrated in Fig.~\ref{fig5}(a), where the topological gap character is plotted as a function of $E_Z$ and $\phi$ for the case of a JJ with $\theta_{so}=\pi/8$ (i.e., $\alpha/\beta\approx 2.4$), $\theta_c=3\pi/4$, and $\varphi=\pi/2$. The vertical dashed line highlights a path along which the JJ transitions between end and edge-like MSs as $\phi$ is varied while keeping $E_Z$ constant. The probability densities of an edge-like MS (cyan triangle) and two end-like MSs (magenta square and magenta dot) are shown in Figs.~\ref{fig5}(b)-(d), where their localization properties are illustrated.

The evolution of the energy spectrum along the path marked with the dashed line in Fig.~\ref{fig5}(a) is displayed in Fig.~\ref{fig5}(e). Red lines represent the energies of the MSs. As the superconducting phase difference is varied, the junction undergoes multiple transitions between end and edge-like MSs. Since all the transitions occur within the TS state, they occur in a protected way, i.e., without gap closings. This is also true for the transitions indicated in Figs.~\ref{fig2}(a), \ref{fig3}(a), and \ref{fig4}(a). It is important to note, however, that for practical applications, the transition path may need to be further optimized with respect to variations in $E_Z$, $\phi$, and system size in order to identify the path between end-like and edge-like MSs that maximizes the topological gap.

\begin{figure}[t]
    \centering
    \includegraphics[width=1\columnwidth]{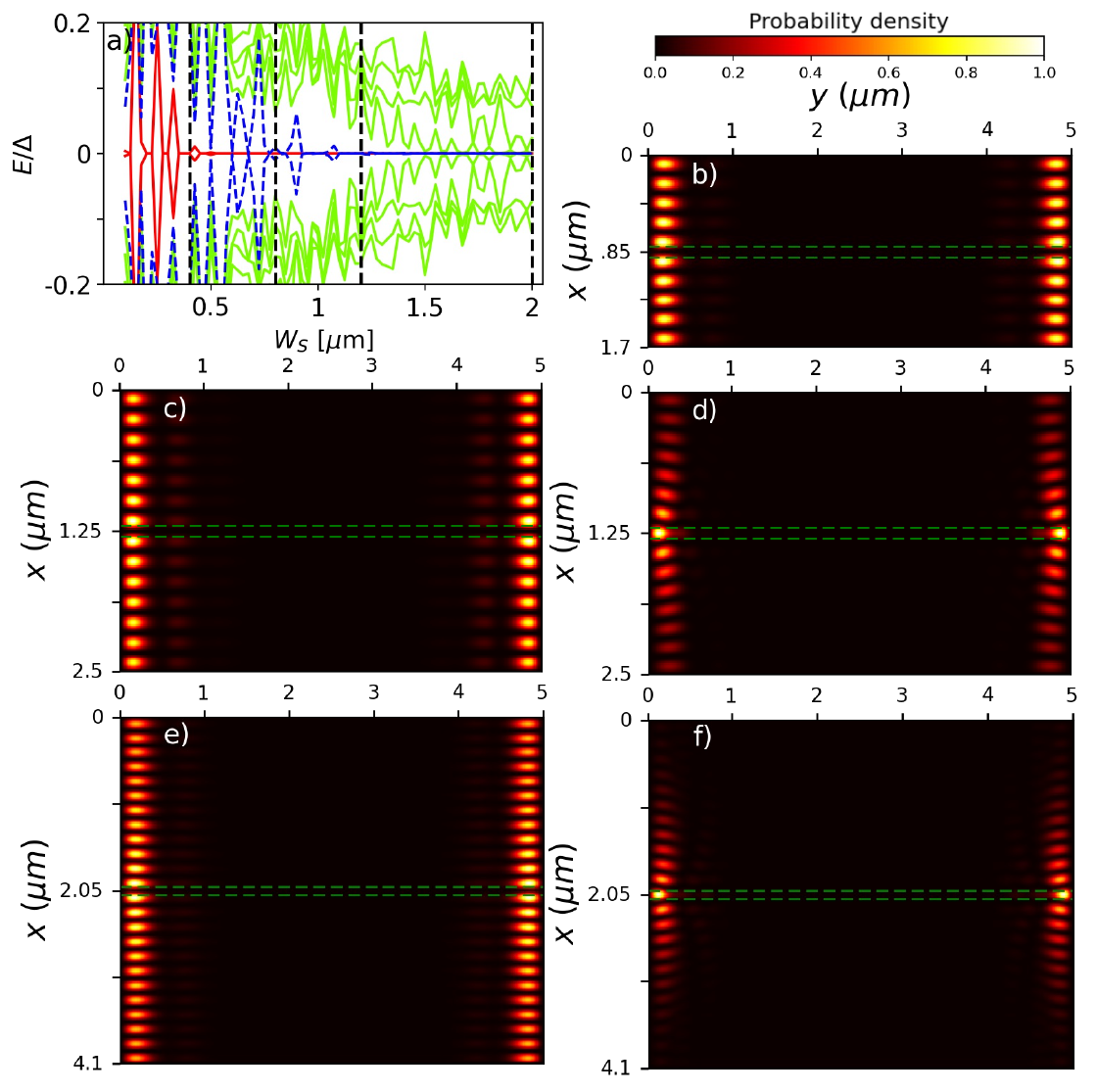}  
    \caption{(a) Energy spectrum as a function of $W_S$ for $\phi=0$ and $E_Z =0.25$~meV. The other system parameters were taken as in Fig.~\ref{fig4}. Red-solid and dashed-blue lines represent states that evolve into MSs as $W_S$
    is varied. Vertical dashed lines indicate $W_S$-values of 0.4~$\mu$m, 0.8~$\mu$m, 1.2~$\mu$m, and 2 $\mu$m.  (b)-(f) Probability density (normalized to its maximum value) for JJs with (b) $W_S=0.8\; \mu$m, (c) and (d) $W_S=1.2\; \mu$m, and (e) and (f) $W_S=2\; \mu$m. The edge-like Majorana pair in (b) evolves into the one shown in (c) (when $W_S$ increases from $W_S=0.8\; \mu$m to $W_S=1.2\; \mu$m) and then into the edge-like pair shown in (e) when $W_S=2\; \mu$m. A second Majorana pair [shown in (d)] exists when $W_S=1.2\; \mu$m and eventually evolves into the end-like MS displayed in (f), which exhibits short localization lengths both along the junction direction and along the system edges perpendicular to the junction}.
    \label{fig6}
\end{figure}

To illustrate the behavior of the MSs as the size of the S regions increases, we show in Fig.~\ref{fig6} the energy spectrum as a function of $W_S$ for $\phi = 0$ and $E_Z =0.25$~meV, with all other system parameters as in Fig.~\ref{fig4}. For clarity, only the 12 states with energies closest to zero are shown. Red-solid and blue-dashed lines indicate states that evolve into MSs as $W_S$ increases. In regions where multiple Majorana pairs coexist, some energy lines may be indistinguishable, as they overlap. Vertical dashed-lines mark the values $W_S=0.4,\;0.8,\;1.2,\;2\;\mu$m.

The probability density (normalized to its maximum value) of the edge-like MSs in a JJ with $W_S=0.4\;\mu$m [see Fig.~\ref{fig4}(b)] transforms into the edge-like MSs shown in Figs.~\ref{fig6}(b), (c), and (e) as $W_S$ increases to 0.8~$\mu$m, 1.2~$\mu$m, and 2~$\mu$m, respectively. A second Majorana pair appears at $W_S=1.2\;\mu$m [see Fig.~\ref{fig6}(d)] and eventually transitions into the end-like Majorana pair displayed in Fig.~\ref{fig6}(f) as the extent of the S regions increase to $2\;\mu$m. Although not depicted in Fig.~\ref{fig6}, a third pair of MSs exists at $W_S=2\;\mu$m.

It is important to note that as long as the chemical potential $\mu$ is larger than the other characteristic energy scales of the system ($\Delta_0$, $E_Z$, and the spin-orbit coupling energy splitting), its specific value has no direct impact on the emergence of edge-like Majorana states (MSs), as suggested by Eq.~(\ref{eq-qvec}). Thus, our calculations for $\mu = 1\;\text{meV}$ (a value larger than the other relevant energy scales) are expected to remain qualitatively accurate even for higher chemical potentials. However, the chemical potential still plays a crucial role, as it affects the extent of the topological region in the $\phi-E_Z$ parameter space. For MSs to emerge, the system must first be driven into the topological superconducting (TS) state. Once in the TS state, the primary effect of the chemical potential is to modulate the oscillations of the edge-like MSs.  

\begin{figure}[t]
    \centering
    \includegraphics[width=27em,height=21em]{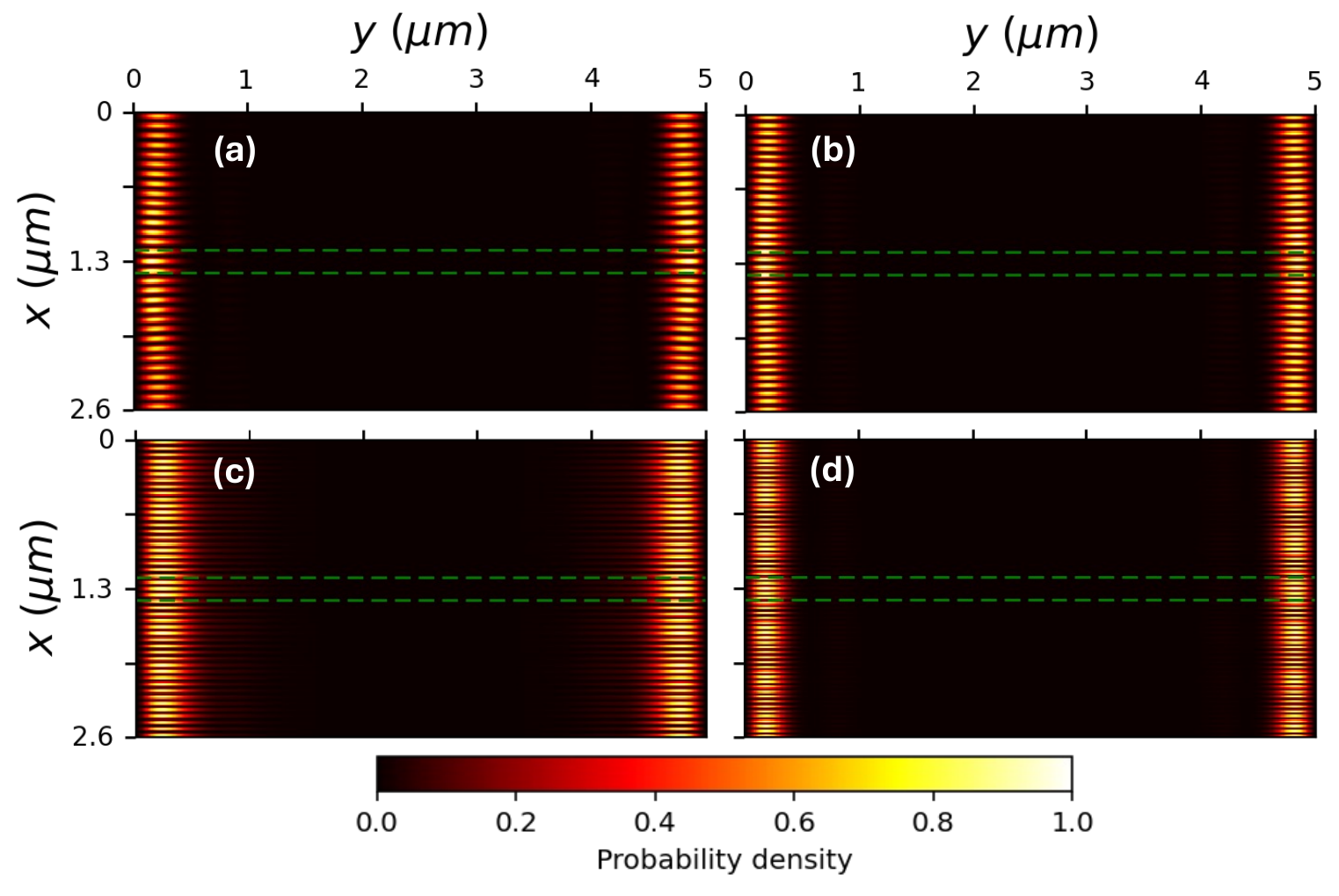}  
    \caption{Probability density (normalized to its maximum value) for JJs with  $W_S=1.2\; \mu$m,  $W_N=0.2\; \mu$m, $E_z=0.25$ meV, and chemical potential (a) $\mu_{S(N)}=4$ meV, (b) $\mu_{S(N)}=6$ meV, (c) $\mu_{S(N)}=12$ meV, and (d) $\mu_{S(N)}=15$ meV. The other parameters were taken as in Fig.~\ref{fig4}(a), namely, $\theta_{so}=\pi/4$, $\theta_c=3\pi/4$, $\phi=0$, and $\varphi_B=\pi/2$.}
    \label{fig7}
\end{figure}

As indicated by Eq.~(\ref{eq-qvec}), when the wave vectors are real, they increase with the chemical potential, leading to shorter wavelengths for the propagating modes and faster oscillations in the probability density of the edge-like MSs. This behavior is evident in Figs.~\ref{fig7}(a)–(d), which display the probability density for junctions with chemical potentials of 4~meV, 6~meV, 12~meV, and 15~meV, respectively.

\begin{figure}[t]
    \centering
    \includegraphics[width=18em,height=22em]{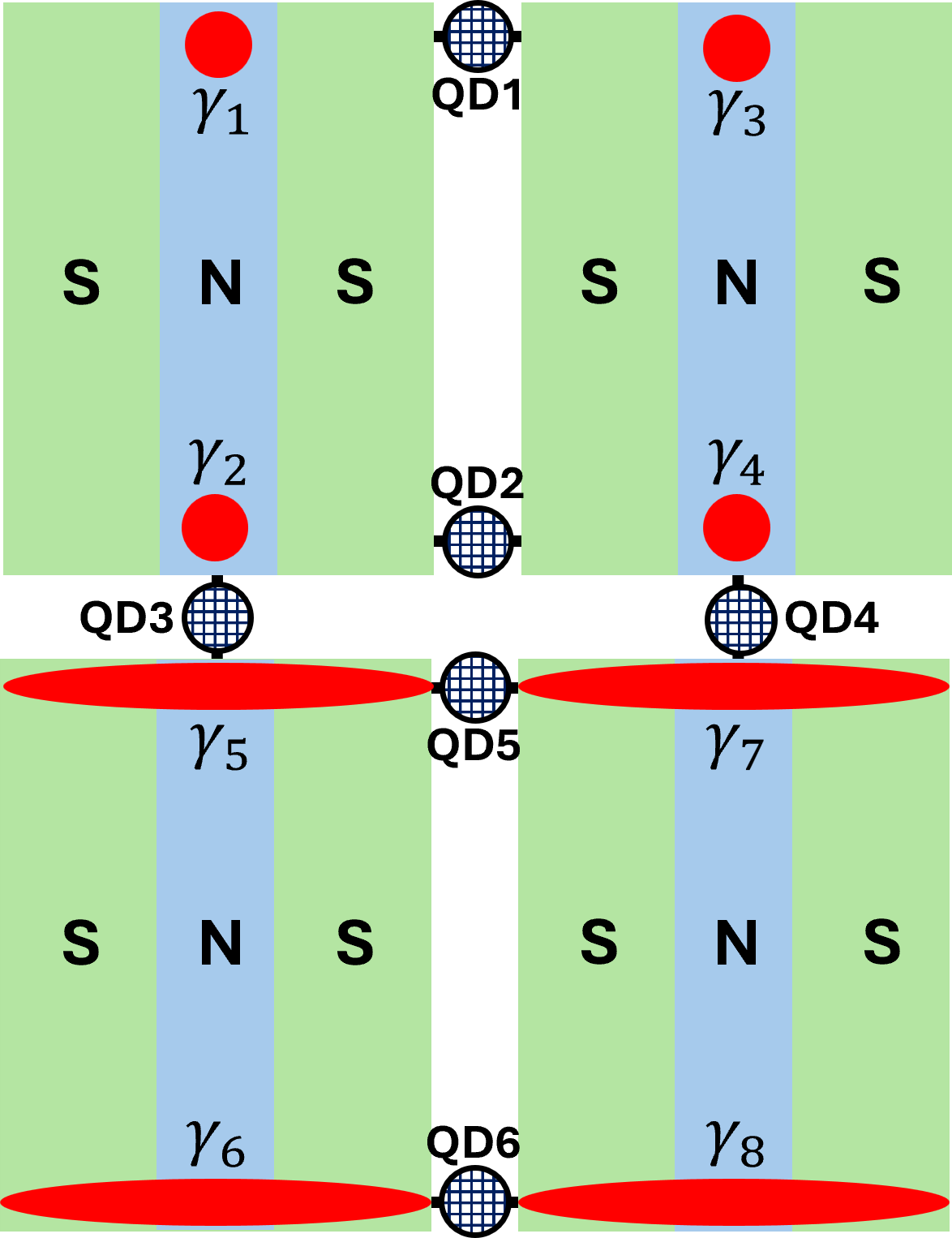}  
    \caption{Schematics of four planar JJs coupled via quantum dots (QDs). Each junction hosts two MSs (red regions), which can be used to encode one topological qubit per junction. The effective coupling between nearby MSs is controlled by QDs QD3 and QD4. Transitions from distant end-like to edge-like MSs enable interconnections between MSs formed in adjacent JJs (e.g., between $\gamma_5$ and $\gamma_7$ or between $\gamma_6$ and $\gamma_8$), where QDs QD1, QD2, QD5, and QD6 can be used to modulate the couplings between MSs along the horizontal direction of the array. Braiding operations can be implemented by appropriately tuning the effective couplings between MSs belonging to different pairs \cite{Flensberg2011:PRL,vanHeck2012:NJP}.}
    \label{fig8}
\end{figure}

Due to their distinct localization characteristics, end-like and edge-like MSs can yield different experimental outcomes. Understanding the behavior and properties of these states is, therefore, crucial for correctly interpreting their experimental detection. For instance, the sharp contrast in the zero-bias conductance peak (ZBCP) observed in Josephson junctions (JJs) when the superconducting phase is changed from  $\phi = 0$ to $\phi = \pi$ \cite{hart2014induced} could be related to a transition between edge-like and end-like MSs. At $\phi = 0$, edge-like MSs exhibit reduced probability density beneath the contact, leading to a smaller ZBCP amplitude. Conversely, at $ \phi = \pi$, end-like MSs have a larger probability density below the contact, resulting in a more prominent ZBCP amplitude. Moreover, because edge-like MSs extend along the junction edges, they could act as global interconnects, enabling coupling between distant MSs. This coupling can be controlled by switching between edge-like and end-like MSs, offering a mechanism for modulating interactions in topological quantum devices and potentially enabling braiding and fusion operations. Figure \ref{fig8} illustrates how transitions to edge-like Majorana states (MSs) can be employed to interconnect MSs formed in adjacent planar Josephson junctions.

\begin{figure}
    \centering
    \includegraphics[width=27em,height=30em]{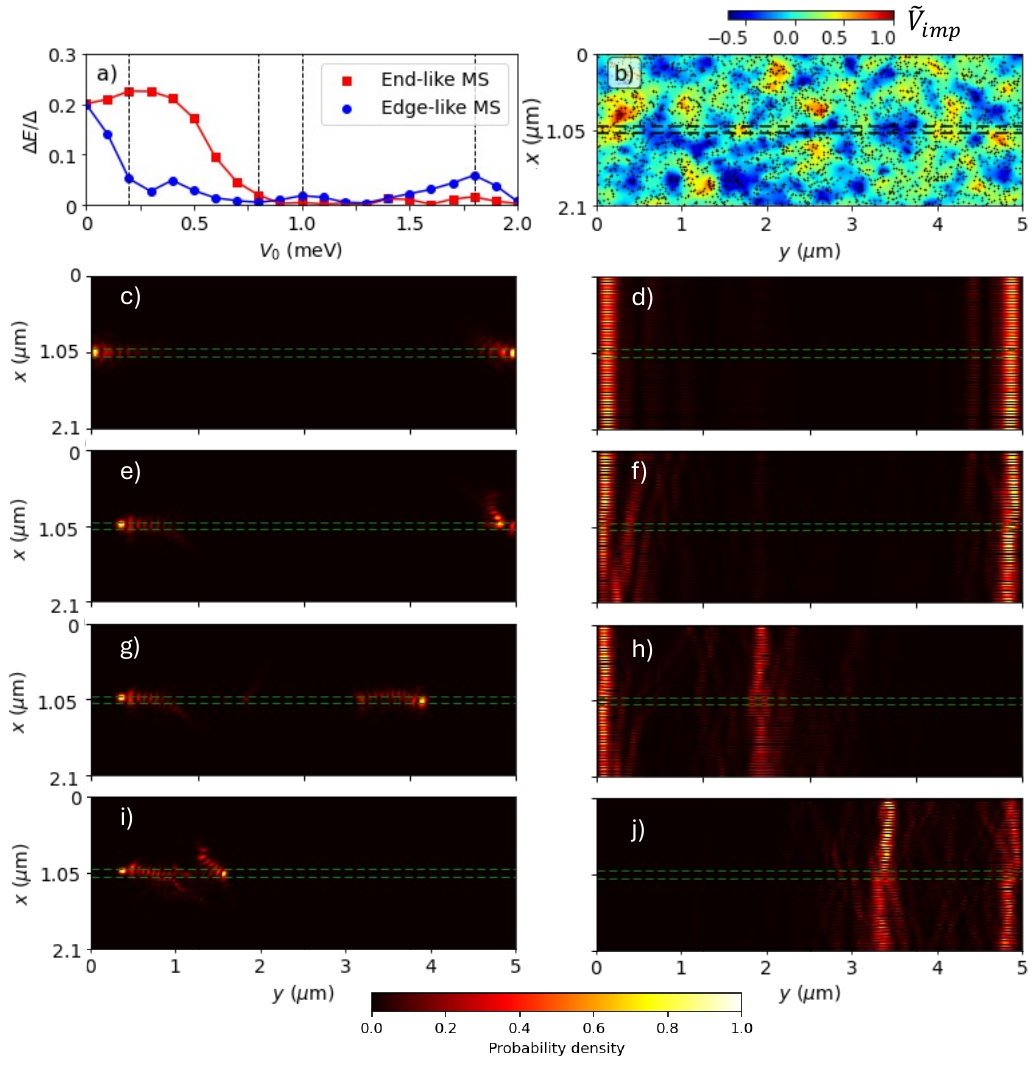}  
    \caption{(a) Difference $\Delta E$ between the two lowest positive energies as a function of the impurity strength $V_0$ for end-like (squares) and edge-like (dots) MSs. (b) Impurity potential landscape normalized to its maximum amplitude $\tilde{V}_{imp}=V_{imp}/max(|V_{imp}|)$. Black dots mark the locations of the impurities. (c)-(j) Probability density (normalized to its maximum value) for end-like (left panel) and edge-like (right panel) MSs. Plots from top to bottom correspond to $V_0=0.2,\;08,\;1.0$, and 1.8~meV [see vertical dashed lines in (a)], respectively. A JJ with $W_N=0.1\;\mu$m, $W_S =1\;\mu$m, and $L=5\;\mu$m and 3\% impurity concentration was considered. For end-like and edge-like MSs we used $E_Z=0.2$~meV [with other parameters as in Fig.~\ref{fig2}(d)] and $E_Z=0.3$~meV [with other parameters as in Fig.~\ref{fig7}(d)], respectively. For these parameters, the end-like and edge-like MSs exhibit the same topological gap ($0.2\Delta$) in the absence of disorder. }
    \label{fig9}
\end{figure}

\subsection{Effects of electrostatic disorder on end-like and edge-like Majorana states}

Randomly distributed impurities in semiconductors introduce fluctuations in the electrostatic landscape of proximitized Josephson junctions (JJs), thereby influencing key properties of the topological superconducting state such as the topological gap and the localization of MSs \cite{Haim2019:PRL,Paudel2205:PReApplied,sharma2024:arxiv,Laubscher2025:PRB,Escribano2025:PRB}. To capture the impact of electrostatic disorder on MSs, we add to the system Hamiltonian [Eq.(\ref{eq1})] an effective disorder potential arising from $N_{imp}$ impurities randomly distributed across the lattice. Following the model described in Ref.~\cite{Paudel2205:PReApplied}, each impurity contributes a local potential with a randomly assigned sign and an amplitude that decays exponentially with distance from the impurity’s position. As a result, the total disorder potential at a given lattice site $\mathbf{r}_i=(x_i,y_i)$ is determined by the cumulative effect of all nearby impurities. The effective disorder potential at the lattice point $\mathbf{r}_i=(x_i,y_i)$ is determined by \cite{Paudel2205:PReApplied},
\begin{equation}\label{eq-vimp}
    V_{imp}(\mathbf{r}_i)=\frac{V_0\left[f(\mathbf{r}_i)-\langle f(\mathbf{r}_i)\rangle\right]}{\sigma_f},
\end{equation}
where $V_0$ is the overall potential amplitude,
\begin{equation}
    f(\mathbf{r}_i)=\sum_{j=1}^{N_{imp}}(-1)^{j}\exp\left(-\frac{|\mathbf{r}_i-\mathbf{R}_j|}{\lambda_{imp}}\right),
\end{equation}
and $\sigma_f$ the standard deviation of $f(\mathbf{r}_i)$. Here $\mathbf{R}_j$ denotes the position of the $j$th impurity and $\lambda_{imp}$ characterizes the decay length of the local impurity potential. The potential defined in Eq.~(\ref{eq-vimp}) satisfies the constraints $\langle V_{imp}\rangle=0$ and $\langle V_{imp}^2\rangle=V_0^2$.

A random sample of $N_{imp}=3171$ impurities in a $(2.1\;\mu{\rm m}\times 5\;\mu{\rm m})$ JJ, corresponding to an impurity concentration of 3\% ($\sim 300\;\mu{\rm m}^{-2}$), and $\lambda_{imp}=70$~nm were used in the numerical simulations. The position dependence of the impurity potential normalized to its maximum amplitude is shown in Fig.~\ref{fig9}(b), where the tiny black dots mark the locations of the randomly distributed impurities.

Figure \ref{fig9}(a) displays the difference $\Delta E$ between the two lowest positive-energy states (which defines the topological gap in the topological superconducting phase) as a function of the disorder potential amplitude $V_0$. Results are shown for both end-like (squares) and edge-like (dots) MSs. Overall, $\Delta E$ decreases with increasing $V_0$, indicating a deterioration of the topological states as disorder becomes stronger. However, this decay is not strictly monotonic and $\Delta E$ exhibits some local maxima, a feature that appears generic. Simulations for multiple disorder realizations (not shown) reveal the same qualitative nonmonotonic behavior, though the precise locations of these maxima vary with the specific disorder configuration.

Interestingly, weak disorder can enhance the topological gap of end-like MSs compared to the clean case. This can be seen in Fig.~\ref{fig9}(a), where the topological gap protecting the end-like MSs is enhanced with respect to the clean cases for finite $V_0$ values up to 0.4~meV.
This observation is consistent with previous studies, which predicted that introducing weak disorder in JJs can enhance the robustness of the topological superconducting phase \cite{Haim2019:PRL}. The topological gap for edge-like MSs tends to decrease more rapidly with $V_0$ in the weak-disorder regime but becomes larger than that for end-like MSs at higher $V_0$. This crossover may not be universal and could shift when the Zeeman field is varied. For clarity, the Zeeman field values were fixed so that the topological gap of end-like and edge-like MSs is the same in the absence of impurities. Nevertheless, at each $V_0$, the gap could likely be further optimized by tuning $E_Z$.

The probability densities of MSs for the $V_0$ values indicated by the vertical dashed lines in Fig.~\ref{fig9} (a) (namely, $V_0=0.2,0.8,1.0$ and 1.8~meV) are displayed in Figs.~\ref{fig9}(c)-(j), where the left and right panels correspond to the end-like and edge-like states, respectively. In the weak-disorder regime, end-like (edge-like) MSs remain well localized near the system ends (edges). However, as the disorder strength increases, the MSs gradually change their localization regions, penetrating deeper into the bulk and eventually approach each other (a similar trend was reported in Ref.~\cite{Paudel2205:PReApplied} for the case of end-like MSs). This behavior poses additional challenges for the practical use of MSs in planar JJs for quantum computing, as maintaining disorder in the weak regime is essential to prevent uncertainties in the MS positions and their reliable manipulation.

\section{Summary}\label{secIV}

We investigated the formation and properties of edge-like and end-like MSs in proximitized planar JJs subjected to an in-plane magnetic field, considering the effects of Rashba and/or Dresselhaus SOCs. The end-like MSs are primarily localized at opposite ends of the normal region within the junction. In contrast, the edge-like MSs extend along the system’s edges, perpendicular to the junction. To characterize the nature and protection of the MSs we introduced a quantity called the \emph{topological gap character}. This quantity provides insight into whether the system is in the D-class (or BDI-class with $\mathbb{Z}_2$ index equal to $\pm 1$) TS state, the size of the topological gap, and whether the MSs are end-like or edge-like. Through numerical simulations of the topological gap character as a function of the magnetic field strength and the superconducting phase difference ($\phi$) across the junction, we found that when the system is in the TS state, and either Rashba or Dresselhaus SOC dominates, edge-like and end-like MSs, protected by a sizable topological gap, typically emerge near $\phi$ values of 0 and $\pi$, respectively. However, when the Rashba and Dresselhaus SOC strengths are comparable, and the junction and magnetic field are properly oriented, protected edge-like MSs can also emerge at phases near $\pi$. Our results reveal the possibility of inducing topologically protected transitions between edge-like and end-like MSs in phase-biased JJs by tuning the superconducting phase difference and the magnetic field strength. Furthermore, in phase-unbiased junctions with comparable Rashba and Dresselhaus SOC strengths, protected transitions between end and edge-like MSs can be achieved by solely adjusting the magnetic field strength, as the superconducting phase self-tunes to minimize the system's free energy. Controlled transitions between edge-like and end-like MSs could function as switchable global interconnects, enabling or disabling the coupling between distant MSs. However, this approach requires maintaining low levels of disorder, since increasing electrostatic disorder can shrink the topological gap and introduce uncertainty in the MS localization.

\section*{Acknowledgments}

A.P.G. is grateful for the funding of PhD scholarship ANID-Chile No. 21210410. P.A.O. acknowledges support from  FONDECYT grants 1230933 and 1220700 and USM-Chile under grant PI-LIR-24-10. A.M.-A. acknowledges support from USM-Chile under grant MEC-USM/2023.

\appendix
\section{}\label{AppendixA}

The system parameters used in the numerical simulations are listed in Table \ref{tab:msg1}. The values of the hopping parameter $t=\hbar^2/(2m^\ast a^2)$ are also included.

\begin{table}[h]                           
 \centering
    \begin{tabular}{|c|c|c|}
    \hline
      parameter & Al/HgTe \cite{Scharf2019:PRB}   &   Al/InSb \cite{mayer2020gate}            
      \\ \hline
     $\Delta_0$ &  0.23 meV  &   0.21 meV             
     \\ \hline          
      $\mu_S$ & 1 meV   &   1 meV
     \\ \hline
     $\mu_N$ & 1 meV   &   1 meV
     \\ \hline
     $m^{*}$ & 0.038$m_o$         &   0.013$m_o$       
    \\ \hline
     $\lambda$ & 16 meV nm  &   15 meV nm
    \\ \hline
     $W_N$ & 100 nm  &   100 nm 
    \\ \hline
    $W_S$ & 400 nm  &   400 nm
    \\ \hline
    $L$ & 5000 nm  &   5000 nm  
     \\ \hline
     $a$ & 10 nm  &   10 nm
    \\ \hline
   $t$ & 10.0 meV  &   29.3 meV
    \\ \hline
    \end{tabular}
    \caption{Material parameters used in the numerical simulations of proximitized planars JJs reported in this work. The proximity-induced superconducting gap is represented by $\Delta_0$, $\mu_S$ ($\mu_N$) denotes the chemical potential in the S (N) region, $m^\ast$  is the electron effective mass (with $m_0$ as the bare mass of the electron), $\lambda=\sqrt{\alpha^2+\beta^2}$ characterizes the combined strength of Rashba ($\alpha$) and Dresselhaus ($\beta$) SOCs, $W_N$ ($W_S$) is the width of the N (S) region, $L$ is the length of the junction, $a$ is the TB lattice constant, and $t$ the TB hopping parameter.}
    \label{tab:msg1}                            
\end{table}

\section{}\label{AppendixB}

To get qualitative insight into the role of the superconducting phase difference in the formation of edge-like and end-like MSs, we assume, without loss of generality that $0\leq\phi\leq\pi$, consider the BdG Hamiltonian in Eq.~(\ref{eq1}), and apply the unitary transformation,
\begin{eqnarray}
    U&=&(\tau_0\otimes\sigma_0)\cos[{\rm sgn}(x)\phi/2]\nonumber\\
    &+&i(\tau_z\otimes\sigma_0)\sin[{\rm sgn}(x)\phi/2]
\end{eqnarray}
which eliminates the phase dependence of superconducting pairing potential at the expense of adding a phase-dependent gauge field to the $ x$ component of the momentum. The transformed Hamiltonian,
\begin{equation}
    H'=U^\dagger HU,
\end{equation}
has the same form as $H$ but with the original pairing potential in Eq.~(\ref{eq6}) replaced by the zero-phase pairing, $\Delta'=\Delta_0\, \Theta(|x|-W_N/2)$, and the momentum component along the x-axis replaced by, $p_x'=p_x \pm (\phi/2)\delta(x)$ in the particle-like (+) and hole-like (-) blocks of the Hamiltonian, respectively. In other words, a JJ with superconducting phase difference $\phi$ is equivalent to a JJ with zero phase ($\phi=0$), but with the Dirac function gauge potentials $A_x=\pm\pm (\phi/2)\delta(x)$ acting on the particle-like and hole-like components of the wavefunction. In the case of a JJ with only Rashba SOC, the gauge potential behaves as an attractive potential in the x-direction. Since it is centered in the middle of the normal (N) region, this potential contributes to the localization of the MSs near the ends of the junction. Conversely, when the gauge potential vanishes for $\phi=0$, the MSs extend along the edges perpendicular to the junction.

Let's now analyze the case in which the Zeeman field dominates the SOC and the spin is approximately conserved. In such a case one can make the approximation $(\mathbf{w}\cdot\mathbf{E}_z)^2\approx (\mathbf{w}\cdot\mathbf{w})^2|\mathbf{E}_z|^2$ in Eq.~(\ref{eq-quartic}) and find approximate analytical solutions for the wave vectors.

\begin{figure}[t]
    \centering
    \includegraphics[width=0.95\columnwidth]{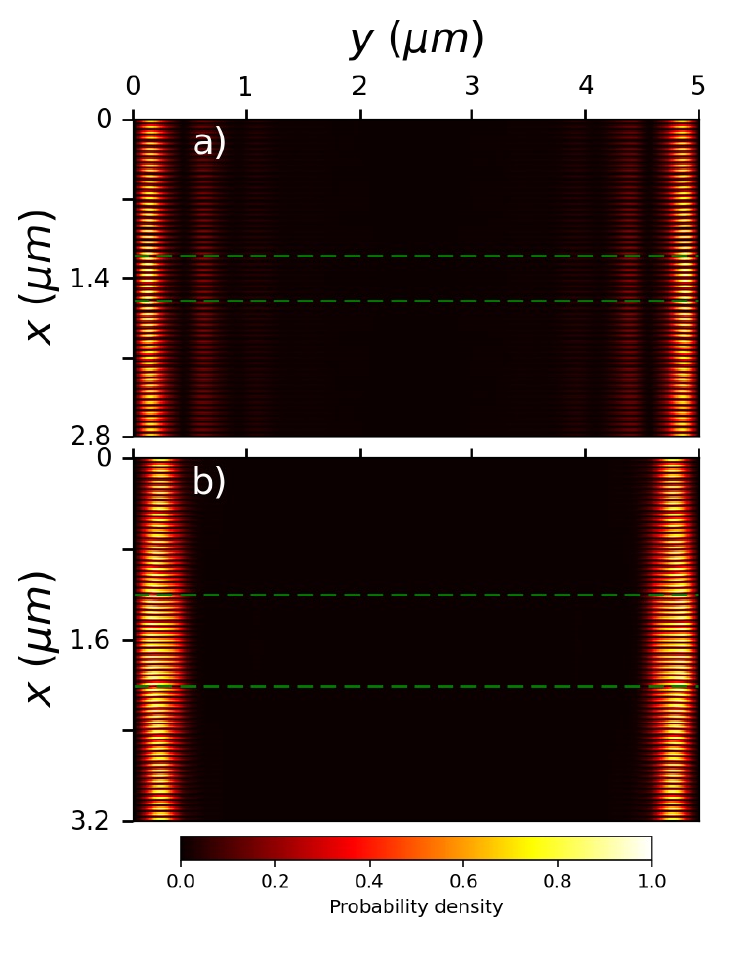} 
    \caption{Probability density (normalized to its maximum value) for JJs with (a) $W_N=0.4\; \mu$m and (b) $W_N=0.8\; \mu$m. The other parameters were taken as in Fig.~\ref{fig7}(c), namely, $W_S=1.2\;\mu$m, $E_Z=0.25$~meV, $\theta_{so}=\pi/4$, $\theta_c=3\pi/4$, $\phi=0$, and $\varphi_B=\pi/2$.}
    \label{fig10}
\end{figure}

\subsection{Magnetic field along the junction ($\mathbf{E}_Z\parallel \hat{\mathbf{y}}$)}

Consider a magnetic field aligned along the junction, together with the absence of Dresselhaus SOC ($\beta=0$) or a junction orientation at an odd multiple of $\pi/4$ with respect to the $[100]$ crystallographic direction of the host semiconductor.
\\
The wave vectors at $\phi=0$ of zero-energy states are given by,
\begin{equation}\label{qeh-1}
     q_{j,\pm}=\sqrt{\left[\dfrac{\sqrt{2m^{*}\left(\mu_S+j\sqrt{E_Z^2 -\Delta_0^2}\right)}}{\hbar} \pm k_{so}\right]^2+\kappa^2}
\end{equation}
where $j=e=1$ ($j=h=-1$) for electron-like (hole-like) states, $E_Z=|\mathbf{E}_Z|$, and
\begin{equation}\label{kso-1}
    k_{so}=\frac{m^\ast(\alpha +\beta \sin 2\theta_c)}{\hbar^2}.
\end{equation}
\\
From Eq.~(\ref{qeh-1},) one finds that the wave vectors are real if,
\begin{equation}\label{g-c-1}
  E_Z\geq \Delta_0\;\;{\rm and}\;\;\mu_S\geq \sqrt{E_Z^2-\Delta_0^2}.
\end{equation}
Hence, edge-like MSs emerge when the system is in the TS state and the conditions in Eq.~(\ref{g-c-1}) are met. Note that, in principle, for the wavefunction to be delocalized, it is sufficient only one, not all, of the wave vectors to be real. In this case, only the first inequality in Eq.~(\ref{g-c-1}) is required. However, when all the wave vectors are real, the wavefunction is guaranteed to be strongly delocalized.

\subsection{Magnetic field perpendicular to the junction ($\mathbf{E}_Z\perp \hat{\mathbf{y}}$)}

In this case, the wave vectors of zero-energy states at $\phi=0$ are found to have the same form as in Eq.~(\ref{qeh-1}) but with,
\begin{equation}\label{kso-1}
    k_{so}=-\frac{m^\ast\beta \cos 2\theta_c}{\hbar^2}.
\end{equation}
\\
We then conclude that in this case, edge-like MSs emerge when the system is in the TS state and the conditions in Eq.~(\ref{g-c-1}) are met.

\section{}\label{AppendixC}

In Sec.~\ref{secIII} we focused our discussion on narrow junctions with $W_N =0.1\;\mu$m [Figs.~\ref{fig3}-\ref{fig6}] and $W_N=0.2\;\mu$m. However, the existence of edge-like MSs is also supported in wider junctions as evidenced by Fig.~\ref{fig10}, where the probability density (normalized to its maximum value) is shown for (a) $W_N=0.4\;\mu$m and (b) $W_N=0.8\;\mu$m. The other system parameters were taken as in Fig.~\ref{fig7}(c).

\twocolumngrid


\bibliographystyle{apsrev4-1}
\bibliography{bibprepint}

@PREAMBLE{
 "\providecommand{\noopsort}[1]{}" 
 # "\providecommand{\singleletter}[1]{#1}%" 
}

@article{Pfaffian,
  author = {Tewari, Sumanta and Sau, Jay D.},
  title = {Topological Invariants for Spin-Orbit Coupled Superconductor Nanowires},
  journal = {Phys. Rev. Lett.},
  volume = {109},
  issue = {15},
  pages = {150408},
  numpages = {5},
  year = {2012},
  month = {Oct},
  publisher = {American Physical Society},
  doi = {10.1103/PhysRevLett.109.150408},
  url = {https://link.aps.org/doi/10.1103/PhysRevLett.109.150408}
}

@article{groth2014kwant,
  author={Groth, Christoph W and Wimmer, Michael and Akhmerov, Anton R and Waintal, Xavier},
  title={Kwant: a software package for quantum transport},
  journal={New Journal of Physics},
  volume={16},
  number={6},
  pages={063065},
  year={2014},
  publisher={IOP Publishing},
  url={https://iopscience.iop.org/article/10.1088/1367-2630/16/6/063065}
}

@article{schnyder2008classification,
  author={Schnyder, Andreas P and Ryu, Shinsei and Furusaki, Akira and Ludwig, Andreas WW},
  title={Classification of topological insulators and superconductors in three spatial dimensions},
  journal={Physical Review B},
  volume={78},
  number={19},
  pages={195125},
  year={2008},
  publisher={APS},
  url={https://journals.aps.org/prb/abstract/10.1103/PhysRevB.78.195125}
}

@article{ryu2010topological,
  author={Ryu, Shinsei and Schnyder, Andreas P and Furusaki, Akira and Ludwig, Andreas WW},
  title={Topological insulators and superconductors: tenfold way and dimensional hierarchy},
  journal={New Journal of Physics},
  volume={12},
  number={6},
  pages={065010},
  year={2010},
  publisher={IOP Publishing},
  url={https://iopscience.iop.org/article/10.1088/1367-2630/12/6/065010/meta}
}

@article{ghosh2010non,
  author={Ghosh, Parag and Sau, Jay D and Tewari, Sumanta and Sarma, S Das},
  title={Non-Abelian topological order in noncentrosymmetric superconductors with broken time-reversal symmetry},
  journal={Physical Review B},
  volume={82},
  number={18},
  pages={184525},
  year={2010},
  publisher={APS},
  url={https://journals.aps.org/prb/abstract/10.1103/PhysRevB.82.184525}
}

@article{rieder2013reentrant,
  author={Rieder, Maria-Theresa and Brouwer, Piet W and Adagideli, Inan{\c{c}}},
  title={Reentrant topological phase transitions in a disordered spinless superconducting wire},
  journal={Physical Review B},
  volume={88},
  number={6},
  pages={060509},
  year={2013},
  publisher={APS},
  url={https://journals.aps.org/prb/abstract/10.1103/PhysRevB.88.060509}
}

@article{adagideli2014effects,
  author={Adagideli, {\.I}nan{\c{c}} and Wimmer, Michael and Teker, Aykut},
  title={Effects of electron scattering on the topological properties of nanowires: Majorana fermions from disorder and superlattices},
  journal={Physical Review B},
  volume={89},
  number={14},
  pages={144506},
  year={2014},
  publisher={APS},
  url={https://journals.aps.org/prb/abstract/10.1103/PhysRevB.89.144506}
}

@article{pekerten2017disorder,
  author={Pekerten, Bar{\i}{\c{s}} and Teker, Aykut and Bozat, {\"O}zg{\"u}r and Wimmer, Michael and Adagideli, {\.I}nan{\c{c}}},
  title={Disorder-induced topological transitions in multichannel Majorana wires},
  journal={Physical Review B},
  volume={95},
  number={6},
  pages={064507},
  year={2017},
  publisher={APS},
  url={https://journals.aps.org/prb/abstract/10.1103/PhysRevB.95.064507}
}

@article{stanescu2011majorana,
  author={Stanescu, Tudor D and Lutchyn, Roman M and Sarma, S Das},
  title={Majorana fermions in semiconductor nanowires},
  journal={Physical Review B},
  volume={84},
  number={14},
  pages={144522},
  year={2011},
  publisher={APS},
  url={https://journals.aps.org/prb/abstract/10.1103/PhysRevB.84.144522}
}

@article{stanescu2013majorana,
  author={Stanescu, Tudor D and Tewari, Sumanta},
  title={Majorana fermions in semiconductor nanowires: fundamentals, modeling, and experiment},
  journal={Journal of Physics: Condensed Matter},
  volume={25},
  number={23},
  pages={233201},
  year={2013},
  publisher={IOP Publishing},
  url={https://iopscience.iop.org/article/10.1088/0953-8984/25/23/233201/meta?casa_token=gZPCgOkhzNkAAAAA:Rt-ReKfXfWMYlDoe16nXfwuXmKZVeVuiPt9RCm_G6N8TVnsjSsEZ59KU1bbe-7dCYwMQFd4fjRtlXj03vA}
}

@article{pekerten2019fermion,
  author={Pekerten, Bar{\i}{\c{s}} and Bozkurt, A Mert and Adagideli, {\.I}nan{\c{c}}},
  title={Fermion parity switches of the ground state of Majorana billiards},
  journal={Physical Review B},
  volume={100},
  number={23},
  pages={235455},
  year={2019},
  publisher={APS},
  url={https://journals.aps.org/prb/abstract/10.1103/PhysRevB.100.235455}
}

@article{kitaev2001unpaired,
  author={Kitaev, A Yu},
  title={Unpaired Majorana fermions in quantum wires},
  journal={Physics-uspekhi},
  volume={44},
  number={10S},
  pages={131},
  year={2001},
  publisher={IOP Publishing},
  url={https://iopscience.iop.org/article/10.1070/1063-7869/44/10S/S29/meta?casa_token=B2_aWk9BxNQAAAAA:RuY46V8ivRLUBfdFCbZ2LQDNMxG_nSwBsUOFGoUZqP5Y_tiE7y5lE6jFhnbNFoLdDqWjZ3kPCyRwrlVjEQ}
}

@article{kitaev2003fault,
  author={Kitaev, A Yu},
  title={Fault-tolerant quantum computation by anyons},
  journal={Annals of physics},
  volume={303},
  number={1},
  pages={2--30},
  year={2003},
  publisher={Elsevier},
  url={https://www.sciencedirect.com/science/article/pii/S0003491602000180?casa_token=2Z7Df8gJWnYAAAAA:y68nYIrZ-amsp47xOvSVCtV6lWr_Ag-rCyQeAvku4_SUaJBbS6ZqSlQEtGyymp26li7PLyFx}
}

@article{qi2011topological,
  author={Qi, Xiao-Liang and Zhang, Shou-Cheng},
  title={Topological insulators and superconductors},
  journal={Reviews of Modern Physics},
  volume={83},
  number={4},
  pages={1057},
  year={2011},
  publisher={APS},
url={https://journals.aps.org/rmp/abstract/10.1103/RevModPhys.83.1057}
}

@article{leijnse2012introduction,
  author={Leijnse, Martin and Flensberg, Karsten},
  title={Introduction to topological superconductivity and Majorana fermions},
  journal={Semiconductor Science and Technology},
  volume={27},
  number={12},
  pages={124003},
  year={2012},
  publisher={IOP Publishing},
url={https://iopscience.iop.org/article/10.1088/0268-1242/27/12/124003}
}

@article{beenakker2013search,
  author={Beenakker, CWJ},
  title={Search for Majorana fermions in superconductors},
  journal={Annu. Rev. Condens. Matter Phys.},
  volume={4},
  number={1},
  pages={113--136},
  year={2013},
  publisher={Annual Reviews},
url={https://www.annualreviews.org/doi/10.1146/annurev-conmatphys-030212-184337}
}

@article{aguado2017majorana,
  author={Aguado, Ram{\'o}n},
  title={Majorana quasiparticles in condensed matter},
  journal={La Rivista del Nuovo Cimento},
  volume={40},
  pages={523--593},
  year={2017},
  publisher={Springer},
url={https://www.sif.it/riviste/sif/ncr/econtents/2017/040/11/article/0}
}

@article{ivanov2001non,
  author={Ivanov, Dmitri A},
  title={Non-Abelian statistics of half-quantum vortices in p-wave superconductors},
  journal={Physical review letters},
  volume={86},
  number={2},
  pages={268},
  year={2001},
  publisher={APS},
url={https://journals.aps.org/prl/abstract/10.1103/PhysRevLett.86.268}
}

@article{nayak2008non,
  author={Nayak, Chetan and Simon, Steven H and Stern, Ady and Freedman, Michael and Sarma, Sankar Das},
  title={Non-Abelian anyons and topological quantum computation},
  journal={Reviews of Modern Physics},
  volume={80},
  number={3},
  pages={1083},
  year={2008},
  publisher={APS},
url={https://journals.aps.org/rmp/abstract/10.1103/RevModPhys.80.1083}
}

@article{alicea2011non,
  author={Alicea, Jason and Oreg, Yuval and Refael, Gil and Von Oppen, Felix and Fisher, Matthew PA},
  title={Non-Abelian statistics and topological quantum information processing in 1D wire networks},
  journal={Nature Physics},
  volume={7},
  number={5},
  pages={412--417},
  year={2011},
  publisher={Nature Publishing Group UK London},
url={https://www.nature.com/articles/nphys1915}
}

@article{aasen2016milestones,
  author={Aasen, David and Hell, Michael and Mishmash, Ryan V and Higginbotham, Andrew and Danon, Jeroen and Leijnse, Martin and Jespersen, Thomas S and Folk, Joshua A and Marcus, Charles M and Flensberg, Karsten and others},
  title={Milestones toward Majorana-based quantum computing},
  journal={Physical Review X},
  volume={6},
  number={3},
  pages={031016},
  year={2016},
  publisher={APS},
url={https://journals.aps.org/prx/abstract/10.1103/PhysRevX.6.031016}
}

@article{Scharf2019:PRB,
  author = {Scharf, Benedikt and Pientka, Falko and Ren, Hechen and Yacoby, Amir and Hankiewicz, Ewelina M.},
  title = {Tuning topological superconductivity in phase-controlled Josephson junctions with Rashba and Dresselhaus spin-orbit coupling},
  journal = {Phys. Rev. B},
  volume = {99},
  issue = {21},
  pages = {214503},
  numpages = {19},
  year = {2019},
  month = {Jun},
  publisher = {American Physical Society},
  doi = {10.1103/PhysRevB.99.214503},
  url = {https://link.aps.org/doi/10.1103/PhysRevB.99.214503}
}

@article{CATSJJ,
  author = {Pakizer, Joseph D. and Scharf, Benedikt and Matos-Abiague, Alex},
  title = {Crystalline anisotropic topological superconductivity in planar Josephson junctions},
  journal = {Phys. Rev. Res.},
  volume = {3},
  issue = {1},
  pages = {013198},
  numpages = {10},
  year = {2021},
  month = {Feb},
  publisher = {American Physical Society},
  doi = {10.1103/PhysRevResearch.3.013198},
  url = {https://link.aps.org/doi/10.1103/PhysRevResearch.3.013198}
}

@article{PhysRevLett.126.036802,
  author = {Dartiailh, Matthieu C. and Mayer, William and Yuan, Joseph and Wickramasinghe, Kaushini S. and Matos-Abiague, Alex and \ifmmode \check{Z}\else \v{Z}\fi{}uti\ifmmode \acute{c}\else \'{c}\fi{}, Igor and Shabani, Javad},
  title = {Phase Signature of Topological Transition in Josephson Junctions},
  journal = {Phys. Rev. Lett.},
  volume = {126},
  issue = {3},
  pages = {036802},
  numpages = {6},
  year = {2021},
  month = {Jan},
  publisher = {American Physical Society},
  doi = {10.1103/PhysRevLett.126.036802},
  url = {https://link.aps.org/doi/10.1103/PhysRevLett.126.036802}
}

@article{mayer2020gate,
  author={Mayer, William and Dartiailh, Matthieu C and Yuan, Joseph and Wickramasinghe, Kaushini S and Rossi, Enrico and Shabani, Javad},
  title={Gate controlled anomalous phase shift in Al/InAs Josephson junctions},
  journal={Nature communications},
  volume={11},
  number={1},
  pages={212},
  year={2020},
  publisher={Nature Publishing Group UK London},
   url={https://www.nature.com/articles/s41467-019-14094-1}
}

@article{PhysRevLett.105.177002,
  author = {Oreg, Yuval and Refael, Gil and von Oppen, Felix},
  title = {Helical Liquids and Majorana Bound States in Quantum Wires},
  journal = {Phys. Rev. Lett.},
  volume = {105},
  issue = {17},
  pages = {177002},
  numpages = {4},
  year = {2010},
  month = {Oct},
  publisher = {American Physical Society},
  doi = {10.1103/PhysRevLett.105.177002},
  url = {https://link.aps.org/doi/10.1103/PhysRevLett.105.177002}
}

@article{PhysRevB.82.214509,
  author = {Sau, Jay D. and Tewari, Sumanta and Lutchyn, Roman M. and Stanescu, Tudor D. and Das Sarma, S.},
  title = {Non-Abelian quantum order in spin-orbit-coupled semiconductors: Search for topological Majorana particles in solid-state systems},
  journal = {Phys. Rev. B},
  volume = {82},
  issue = {21},
  pages = {214509},
  numpages = {26},
  year = {2010},
  month = {Dec},
  publisher = {American Physical Society},
  doi = {10.1103/PhysRevB.82.214509},
  url = {https://link.aps.org/doi/10.1103/PhysRevB.82.214509}
}

@article{PhysRevLett.104.040502,
  author = {Sau, Jay D. and Lutchyn, Roman M. and Tewari, Sumanta and Das Sarma, S.},
  title = {Generic New Platform for Topological Quantum Computation Using Semiconductor Heterostructures},
  journal = {Phys. Rev. Lett.},
  volume = {104},
  issue = {4},
  pages = {040502},
  numpages = {4},
  year = {2010},
  month = {Jan},
  publisher = {American Physical Society},
  doi = {10.1103/PhysRevLett.104.040502},
  url = {https://link.aps.org/doi/10.1103/PhysRevLett.104.040502}
}

@article{PhysRevLett.105.077001,
  author = {Lutchyn, Roman M. and Sau, Jay D. and Das Sarma, S.},
  title = {Majorana Fermions and a Topological Phase Transition in Semiconductor-Superconductor Heterostructures},
  journal = {Phys. Rev. Lett.},
  volume = {105},
  issue = {7},
  pages = {077001},
  numpages = {4},
  year = {2010},
  month = {Aug},
  publisher = {American Physical Society},
  doi = {10.1103/PhysRevLett.105.077001},
  url = {https://link.aps.org/doi/10.1103/PhysRevLett.105.077001}
}

@article{rokhinson2012fractional,
  author={Rokhinson, Leonid P and Liu, Xinyu and Furdyna, Jacek K},
  title={The fractional ac Josephson effect in a semiconductor--superconductor nanowire as a signature of Majorana particles},
  journal={Nature Physics},
  volume={8},
  number={11},
  pages={795--799},
  year={2012},
  publisher={Nature Publishing Group UK London},
url={https://www.nature.com/articles/nphys2429#citeas}
}

@article{PhysRevLett.109.227006,
  author = {Pientka, Falko and Kells, Graham and Romito, Alessandro and Brouwer, Piet W. and von Oppen, Felix},
  title = {Enhanced Zero-Bias Majorana Peak in the Differential Tunneling Conductance of Disordered Multisubband Quantum-Wire/Superconductor Junctions},
  journal = {Phys. Rev. Lett.},
  volume = {109},
  issue = {22},
  pages = {227006},
  numpages = {5},
  year = {2012},
  month = {Nov},
  publisher = {American Physical Society},
  doi = {10.1103/PhysRevLett.109.227006},
  url = {https://link.aps.org/doi/10.1103/PhysRevLett.109.227006}
}

@article{mourik2012signatures,
  author={Mourik, Vincent and Zuo, Kun and Frolov, Sergey M and Plissard, SR and Bakkers, Erik PAM and Kouwenhoven, Leo P},
  title={Signatures of Majorana fermions in hybrid superconductor-semiconductor nanowire devices},
  journal={Science},
  volume={336},
  number={6084},
  pages={1003--1007},
  year={2012},
  publisher={American Association for the Advancement of Science},
url={https://www.science.org/doi/10.1126/science.1222360}
}

@article{das2012zero,
  author={Das, Anindya and Ronen, Yuval and Most, Yonatan and Oreg, Yuval and Heiblum, Moty and Shtrikman, Hadas},
  title={Zero-bias peaks and splitting in an Al--InAs nanowire topological superconductor as a signature of Majorana fermions},
  journal={Nature Physics},
  volume={8},
  number={12},
  pages={887--895},
  year={2012},
  publisher={Nature Publishing Group UK London},
url={https://www.nature.com/articles/nphys2479}
}

@article{deng2012anomalous,
  author={Deng, MT and Yu, CL and Huang, GY and Larsson, Marcus and Caroff, Philippe and Xu, HQ},
  title={Anomalous zero-bias conductance peak in a Nb--InSb nanowire--Nb hybrid device},
  journal={Nano letters},
  volume={12},
  number={12},
  pages={6414--6419},
  year={2012},
  publisher={ACS Publications},
url={https://pubs.acs.org/doi/10.1021/nl303758w}
}

@article{deng2016majorana,
  author={Deng, MT and Vaitiek{\.e}nas, S and Hansen, Esben Bork and Danon, Jeroen and Leijnse, M and Flensberg, Karsten and Nyg{\aa}rd, Jesper and Krogstrup, P and Marcus, Charles M},
  title={Majorana bound state in a coupled quantum-dot hybrid-nanowire system},
  journal={Science},
  volume={354},
  number={6319},
  pages={1557--1562},
  year={2016},
  publisher={American Association for the Advancement of Science},
url={https://www.science.org/doi/10.1126/science.aaf3961}
}

@article{manna2020signature,
  author={Manna, Sujit and Wei, Peng and Xie, Yingming and Law, Kam Tuen and Lee, Patrick A and Moodera, Jagadeesh S},
  title={Signature of a pair of Majorana zero modes in superconducting gold surface states},
  journal={Proceedings of the National Academy of Sciences},
  volume={117},
  number={16},
  pages={8775--8782},
  year={2020},
  publisher={National Acad Sciences},
url={https://www.pnas.org/doi/full/10.1073/pnas.1919753117}
}

@article{PhysRevB.84.195442,
  author = {Choy, T.-P. and Edge, J. M. and Akhmerov, A. R. and Beenakker, C. W. J.},
  title = {Majorana fermions emerging from magnetic nanoparticles on a superconductor without spin-orbit coupling},
  journal = {Phys. Rev. B},
  volume = {84},
  issue = {19},
  pages = {195442},
  numpages = {6},
  year = {2011},
  month = {Nov},
  publisher = {American Physical Society},
  doi = {10.1103/PhysRevB.84.195442},
  url = {https://link.aps.org/doi/10.1103/PhysRevB.84.195442}
}

@article{PhysRevB.85.144505,
  author = {Martin, Ivar and Morpurgo, Alberto F.},
  title = {Majorana fermions in superconducting helical magnets},
  journal = {Phys. Rev. B},
  volume = {85},
  issue = {14},
  pages = {144505},
  numpages = {6},
  year = {2012},
  month = {Apr},
  publisher = {American Physical Society},
  doi = {10.1103/PhysRevB.85.144505},
  url = {https://link.aps.org/doi/10.1103/PhysRevB.85.144505}
}

@article{PhysRevB.88.155420,
  author = {Pientka, Falko and Glazman, Leonid I. and von Oppen, Felix},
  title = {Topological superconducting phase in helical Shiba chains},
  journal = {Phys. Rev. B},
  volume = {88},
  issue = {15},
  pages = {155420},
  numpages = {13},
  year = {2013},
  month = {Oct},
  publisher = {American Physical Society},
  doi = {10.1103/PhysRevB.88.155420},
  url = {https://link.aps.org/doi/10.1103/PhysRevB.88.155420}
}

@article{PhysRevB.88.020407,
  author = {Nadj-Perge, S. and Drozdov, I. K. and Bernevig, B. A. and Yazdani, Ali},
  title = {Proposal for realizing Majorana fermions in chains of magnetic atoms on a superconductor},
  journal = {Phys. Rev. B},
  volume = {88},
  issue = {2},
  pages = {020407},
  numpages = {5},
  year = {2013},
  month = {Jul},
  publisher = {American Physical Society},
  doi = {10.1103/PhysRevB.88.020407},
  url = {https://link.aps.org/doi/10.1103/PhysRevB.88.020407}
}

@article{nadj2014observation,
  author={Nadj-Perge, Stevan and Drozdov, Ilya K and Li, Jian and Chen, Hua and Jeon, Sangjun and Seo, Jungpil and MacDonald, Allan H and Bernevig, B Andrei and Yazdani, Ali},
  title={Observation of Majorana fermions in ferromagnetic atomic chains on a superconductor},
  journal={Science},
  volume={346},
  number={6209},
  pages={602--607},
  year={2014},
  publisher={American Association for the Advancement of Science},
url={https://www.science.org/doi/10.1126/science.1259327}
}

@article{pawlak2016probing,
  author={Pawlak, R{\'e}my and Kisiel, Marcin and Klinovaja, Jelena and Meier, Tobias and Kawai, Shigeki and Glatzel, Thilo and Loss, Daniel and Meyer, Ernst},
  title={Probing atomic structure and Majorana wavefunctions in mono-atomic Fe chains on superconducting Pb surface},
  journal={npj Quantum Information},
  volume={2},
  number={1},
  pages={1--5},
  year={2016},
  publisher={Nature Publishing Group},
url={https://www.nature.com/articles/npjqi201635}
}

@article{PhysRevLett.117.077002,
  author = {Fatin, Geoffrey L. and Matos-Abiague, Alex and Scharf, Benedikt and \ifmmode \check{Z}\else \v{Z}\fi{}uti\ifmmode \acute{c}\else \'{c}\fi{}, Igor},
  title = {Wireless Majorana Bound States: From Magnetic Tunability to Braiding},
  journal = {Phys. Rev. Lett.},
  volume = {117},
  issue = {7},
  pages = {077002},
  numpages = {6},
  year = {2016},
  month = {Aug},
  publisher = {American Physical Society},
  doi = {10.1103/PhysRevLett.117.077002},
  url = {https://link.aps.org/doi/10.1103/PhysRevLett.117.077002}
}

@article{matos2017tunable,
  author={Matos-Abiague, Alex and Shabani, Javad and Kent, Andrew D and Fatin, Geoffrey L and Scharf, Benedikt and {\v{Z}}uti{\'c}, Igor},
  title={Tunable magnetic textures: From Majorana bound states to braiding},
  journal={Solid State Communications},
  volume={262},
  pages={1--6},
  year={2017},
  publisher={Elsevier},
url={https://www.sciencedirect.com/science/article/abs/pii/S0038109817301904?via%3Dihub}
}

@article{PhysRevApplied.12.034048,
  author = {Mohanta, Narayan and Zhou, Tong and Xu, Jun-Wen and Han, Jong E. and Kent, Andrew D. and Shabani, Javad and \ifmmode \check{Z}\else \v{Z}\fi{}uti\ifmmode \acute{c}\else \'{c}\fi{}, Igor and Matos-Abiague, Alex},
  title = {Electrical Control of Majorana Bound States Using Magnetic Stripes},
  journal = {Phys. Rev. Appl.},
  volume = {12},
  issue = {3},
  pages = {034048},
  numpages = {9},
  year = {2019},
  month = {Sep},
  publisher = {American Physical Society},
  doi = {10.1103/PhysRevApplied.12.034048},
  url = {https://link.aps.org/doi/10.1103/PhysRevApplied.12.034048}
}

@article{PhysRevB.99.134505,
  author = {Zhou, Tong and Mohanta, Narayan and Han, Jong E. and Matos-Abiague, Alex and \ifmmode \check{Z}\else \v{Z}\fi{}uti\ifmmode \acute{c}\else \'{c}\fi{}, Igor},
  title = {Tunable magnetic textures in spin valves: From spintronics to Majorana bound states},
  journal = {Phys. Rev. B},
  volume = {99},
  issue = {13},
  pages = {134505},
  numpages = {8},
  year = {2019},
  month = {Apr},
  publisher = {American Physical Society},
  doi = {10.1103/PhysRevB.99.134505},
  url = {https://link.aps.org/doi/10.1103/PhysRevB.99.134505}
}

@article{PhysRevLett.109.236801,
  author = {Klinovaja, Jelena and Stano, Peter and Loss, Daniel},
  title = {Transition from Fractional to Majorana Fermions in Rashba Nanowires},
  journal = {Phys. Rev. Lett.},
  volume = {109},
  issue = {23},
  pages = {236801},
  numpages = {5},
  year = {2012},
  month = {Dec},
  publisher = {American Physical Society},
  doi = {10.1103/PhysRevLett.109.236801},
  url = {https://link.aps.org/doi/10.1103/PhysRevLett.109.236801}
}

@article{PhysRevB.85.020503,
  author = {Kjaergaard, Morten and W\"olms, Konrad and Flensberg, Karsten},
  title = {Majorana fermions in superconducting nanowires without spin-orbit coupling},
  journal = {Phys. Rev. B},
  volume = {85},
  issue = {2},
  pages = {020503},
  numpages = {4},
  year = {2012},
  month = {Jan},
  publisher = {American Physical Society},
  doi = {10.1103/PhysRevB.85.020503},
  url = {https://link.aps.org/doi/10.1103/PhysRevB.85.020503}
}

@article{PhysRevB.95.140504,
  author = {Marra, Pasquale and Cuoco, Mario},
  title = {Controlling Majorana states in topologically inhomogeneous superconductors},
  journal = {Phys. Rev. B},
  volume = {95},
  issue = {14},
  pages = {140504},
  numpages = {6},
  year = {2017},
  month = {Apr},
  publisher = {American Physical Society},
  doi = {10.1103/PhysRevB.95.140504},
  url = {https://link.aps.org/doi/10.1103/PhysRevB.95.140504}
}

@article{desjardins2019synthetic,
  author={Desjardins, MM and Contamin, LC and Delbecq, MR and Dartiailh, MC and Bruhat, LE and Cubaynes, T and Viennot, JJ and Mallet, F and Rohart, S and Thiaville, A and others},
  title={Synthetic spin--orbit interaction for Majorana devices},
  journal={Nature materials},
  volume={18},
  number={10},
  pages={1060--1064},
  year={2019},
  publisher={Nature Publishing Group UK London},
url={https://www.nature.com/articles/s41563-019-0457-6}
}

@article{PhysRevB.104.174502,
  author = {Steffensen, Daniel and Andersen, Brian M. and Kotetes, Panagiotis},
  title = {Trapping Majorana zero modes in vortices of magnetic texture crystals coupled to nodal superconductors},
  journal = {Phys. Rev. B},
  volume = {104},
  issue = {17},
  pages = {174502},
  numpages = {16},
  year = {2021},
  month = {Nov},
  publisher = {American Physical Society},
  doi = {10.1103/PhysRevB.104.174502},
  url = {https://link.aps.org/doi/10.1103/PhysRevB.104.174502}
}

@article{Pientka2017:PRX,
  author = {Pientka, Falko and Keselman, Anna and Berg, Erez and Yacoby, Amir and Stern, Ady and Halperin, Bertrand I.},
  title = {Topological Superconductivity in a Planar Josephson Junction},
  journal = {Phys. Rev. X},
  volume = {7},
  issue = {2},
  pages = {021032},
  numpages = {17},
  year = {2017},
  month = {May},
  publisher = {American Physical Society},
  doi = {10.1103/PhysRevX.7.021032},
  url = {https://link.aps.org/doi/10.1103/PhysRevX.7.021032}
}

@article{Setiawan2019:PRBb,
  author = {Setiawan, F. and Stern, Ady and Berg, Erez},
  title = {Topological superconductivity in planar Josephson junctions: Narrowing down to the nanowire limit},
  journal = {Phys. Rev. B},
  volume = {99},
  issue = {22},
  pages = {220506},
  numpages = {7},
  year = {2019},
  month = {Jun},
  publisher = {American Physical Society},
  doi = {10.1103/PhysRevB.99.220506},
  url = {https://link.aps.org/doi/10.1103/PhysRevB.99.220506}
}

@article{Setiawan2019:PRBa,
  author = {Setiawan, F. and Wu, Chien-Te and Levin, K.},
  title = {Full proximity treatment of topological superconductors in Josephson-junction architectures},
  journal = {Phys. Rev. B},
  volume = {99},
  issue = {17},
  pages = {174511},
  numpages = {16},
  year = {2019},
  month = {May},
  publisher = {American Physical Society},
  doi = {10.1103/PhysRevB.99.174511},
  url = {https://link.aps.org/doi/10.1103/PhysRevB.99.174511}
}

@article{fornieri2019evidence,
  author={Fornieri, Antonio and Whiticar, Alexander M and Setiawan, F and Portol{\'e}s, El{\'\i}as and Drachmann, Asbj{\o}rn CC and Keselman, Anna and Gronin, Sergei and Thomas, Candice and Wang, Tian and Kallaher, Ray and others},
  title={Evidence of topological superconductivity in planar Josephson junctions},
  journal={Nature},
  volume={569},
  number={7754},
  pages={89--92},
  year={2019},
  publisher={Nature Publishing Group UK London},
url={https://www.nature.com/articles/s41586-019-1068-8}
}

@article{ren2019topological,
  author={Ren, Hechen and Pientka, Falko and Hart, Sean and Pierce, Andrew T and Kosowsky, Michael and Lunczer, Lukas and Schlereth, Raimund and Scharf, Benedikt and Hankiewicz, Ewelina M and Molenkamp, Laurens W and others},
  title={Topological superconductivity in a phase-controlled Josephson junction},
  journal={Nature},
  volume={569},
  number={7754},
  pages={93--98},
  year={2019},
  publisher={Nature Publishing Group UK London},
url={https://www.nature.com/articles/s41586-019-1148-9}
}

@article{hart2014induced,
  author={Hart, Sean and Ren, Hechen and Wagner, Timo and Leubner, Philipp and M{\"u}hlbauer, Mathias and Br{\"u}ne, Christoph and Buhmann, Hartmut and Molenkamp, Laurens W and Yacoby, Amir},
  title={Induced superconductivity in the quantum spin Hall edge},
  journal={Nature Physics},
  volume={10},
  number={9},
  pages={638--643},
  year={2014},
  publisher={Nature Publishing Group UK London},
url={https://www.nature.com/articles/nphys3036}
}

@article{PhysRevLett.125.086802,
  author = {Laeven, Tom and Nijholt, Bas and Wimmer, Michael and Akhmerov, Anton R.},
  title = {Enhanced Proximity Effect in Zigzag-Shaped Majorana Josephson Junctions},
  journal = {Phys. Rev. Lett.},
  volume = {125},
  issue = {8},
  pages = {086802},
  numpages = {5},
  year = {2020},
  month = {Aug},
  publisher = {American Physical Society},
  doi = {10.1103/PhysRevLett.125.086802},
  url = {https://link.aps.org/doi/10.1103/PhysRevLett.125.086802}
}

@article{lesser2021phase,
  author={Lesser, Omri and Saydjari, Andrew and Wesson, Marie and Yacoby, Amir and Oreg, Yuval},
  title={Phase-induced topological superconductivity in a planar heterostructure},
  journal={Proceedings of the National Academy of Sciences},
  volume={118},
  number={27},
  pages={e2107377118},
  year={2021},
  publisher={National Acad Sciences},
url={https://www.pnas.org/doi/full/10.1073/pnas.2107377118}
}

@article{zhou2022fusion,
  author={Zhou, Tong and Dartiailh, Matthieu C and Sardashti, Kasra and Han, Jong E and Matos-Abiague, Alex and Shabani, Javad and {\v{Z}}uti{\'c}, Igor},
  title={Fusion of Majorana bound states with mini-gate control in two-dimensional systems},
  journal={Nature Communications},
  volume={13},
  number={1},
  pages={1738},
  year={2022},
  publisher={Nature Publishing Group UK London},
url={https://www.nature.com/articles/s41467-022-29463-6}
}

@article{PhysRevLett.124.137001,
  author = {Zhou, Tong and Dartiailh, Matthieu C. and Mayer, William and Han, Jong E. and Matos-Abiague, Alex and Shabani, Javad and \ifmmode \check{Z}\else \v{Z}\fi{}uti\ifmmode \acute{c}\else \'{c}\fi{}, Igor},
  title = {Phase Control of Majorana Bound States in a Topological $\mathsf{X}$ Junction},
  journal = {Phys. Rev. Lett.},
  volume = {124},
  issue = {13},
  pages = {137001},
  numpages = {7},
  year = {2020},
  month = {Apr},
  publisher = {American Physical Society},
  doi = {10.1103/PhysRevLett.124.137001},
  url = {https://link.aps.org/doi/10.1103/PhysRevLett.124.137001}
}

@article{PhysRevB.99.214503,
  author = {Scharf, Benedikt and Pientka, Falko and Ren, Hechen and Yacoby, Amir and Hankiewicz, Ewelina M.},
  title = {Tuning topological superconductivity in phase-controlled Josephson junctions with Rashba and Dresselhaus spin-orbit coupling},
  journal = {Phys. Rev. B},
  volume = {99},
  issue = {21},
  pages = {214503},
  numpages = {19},
  year = {2019},
  month = {Jun},
  publisher = {American Physical Society},
  doi = {10.1103/PhysRevB.99.214503},
  url = {https://link.aps.org/doi/10.1103/PhysRevB.99.214503}
}

@article{PhysRevB.96.205425,
  author = {Cayao, Jorge and San-Jose, Pablo and Black-Schaffer, Annica M. and Aguado, Ram\'on and Prada, Elsa},
  title = {Majorana splitting from critical currents in Josephson junctions},
  journal = {Phys. Rev. B},
  volume = {96},
  issue = {20},
  pages = {205425},
  numpages = {9},
  year = {2017},
  month = {Nov},
  publisher = {American Physical Society},
  doi = {10.1103/PhysRevB.96.205425},
  url = {https://link.aps.org/doi/10.1103/PhysRevB.96.205425}
}

@article{PhysRevLett.89.137007,
  author = {Kontos, T. and Aprili, M. and Lesueur, J. and Gen\^et, F. and Stephanidis, B. and Boursier, R.},
  title = {Josephson Junction through a Thin Ferromagnetic Layer: Negative Coupling},
  journal = {Phys. Rev. Lett.},
  volume = {89},
  issue = {13},
  pages = {137007},
  numpages = {4},
  year = {2002},
  month = {Sep},
  publisher = {American Physical Society},
  doi = {10.1103/PhysRevLett.89.137007},
  url = {https://link.aps.org/doi/10.1103/PhysRevLett.89.137007}
}

@article{PhysRevB.89.195407,
  author = {Yokoyama, Tomohiro and Eto, Mikio and Nazarov, Yuli V.},
  title = {Anomalous Josephson effect induced by spin-orbit interaction and Zeeman effect in semiconductor nanowires},
  journal = {Phys. Rev. B},
  volume = {89},
  issue = {19},
  pages = {195407},
  numpages = {14},
  year = {2014},
  month = {May},
  publisher = {American Physical Society},
  doi = {10.1103/PhysRevB.89.195407},
  url = {https://link.aps.org/doi/10.1103/PhysRevB.89.195407}
}

@article{PhysRevLett.118.107701,
  author = {Hell, Michael and Leijnse, Martin and Flensberg, Karsten},
  title = {Two-Dimensional Platform for Networks of Majorana Bound States},
  journal = {Phys. Rev. Lett.},
  volume = {118},
  issue = {10},
  pages = {107701},
  numpages = {6},
  year = {2017},
  month = {Mar},
  publisher = {American Physical Society},
  doi = {10.1103/PhysRevLett.118.107701},
  url = {https://link.aps.org/doi/10.1103/PhysRevLett.118.107701}
}

@article{PhysRevB.102.245403,
  author = {Zhang, Yu and Guo, Kaige and Liu, Jie},
  title = {Transport characterization of topological superconductivity in a planar Josephson junction},
  journal = {Phys. Rev. B},
  volume = {102},
  issue = {24},
  pages = {245403},
  numpages = {8},
  year = {2020},
  month = {Dec},
  publisher = {American Physical Society},
  doi = {10.1103/PhysRevB.102.245403},
  url = {https://link.aps.org/doi/10.1103/PhysRevB.102.245403}
}

@article{PhysRevB.101.195435,
  author = {Woods, Benjamin D. and Stanescu, Tudor D.},
  title = {Enhanced topological protection in planar quasi-one-dimensional channels with periodically modulated width},
  journal = {Phys. Rev. B},
  volume = {101},
  issue = {19},
  pages = {195435},
  numpages = {18},
  year = {2020},
  month = {May},
  publisher = {American Physical Society},
  doi = {10.1103/PhysRevB.101.195435},
  url = {https://link.aps.org/doi/10.1103/PhysRevB.101.195435}
}

@article{PhysRevB.99.035307,
  author = {Stenger, John P. T. and Hatridge, Michael and Frolov, Sergey M. and Pekker, David},
  title = {Braiding quantum circuit based on the $4\ensuremath{\pi}$ Josephson effect},
  journal = {Phys. Rev. B},
  volume = {99},
  issue = {3},
  pages = {035307},
  numpages = {9},
  year = {2019},
  month = {Jan},
  publisher = {American Physical Society},
  doi = {10.1103/PhysRevB.99.035307},
  url = {https://link.aps.org/doi/10.1103/PhysRevB.99.035307}
}

@article{PhysRevB.103.L180505,
  author = {Svetogorov, Aleksandr E. and Loss, Daniel and Klinovaja, Jelena},
  title = {Insulating regime of an underdamped current-biased Josephson junction supporting ${\mathbb{Z}}_{3}$ and ${\mathbb{Z}}_{4}$ parafermions},
  journal = {Phys. Rev. B},
  volume = {103},
  issue = {18},
  pages = {L180505},
  numpages = {5},
  year = {2021},
  month = {May},
  publisher = {American Physical Society},
  doi = {10.1103/PhysRevB.103.L180505},
  url = {https://link.aps.org/doi/10.1103/PhysRevB.103.L180505}
}

@article{PhysRevLett.111.107007,
  author = {Pekker, David and Hou, Chang-Yu and Manucharyan, Vladimir E. and Demler, Eugene},
  title = {Proposal for Coherent Coupling of Majorana Zero Modes and Superconducting Qubits Using the $4\ensuremath{\pi}$ Josephson Effect},
  journal = {Phys. Rev. Lett.},
  volume = {111},
  issue = {10},
  pages = {107007},
  numpages = {5},
  year = {2013},
  month = {Sep},
  publisher = {American Physical Society},
  doi = {10.1103/PhysRevLett.111.107007},
  url = {https://link.aps.org/doi/10.1103/PhysRevLett.111.107007}
}

@article{PhysRevB.104.155428,
  author = {Paudel, Purna P. and Cole, Trey and Woods, Benjamin D. and Stanescu, Tudor D.},
  title = {Enhanced topological superconductivity in spatially modulated planar Josephson junctions},
  journal = {Phys. Rev. B},
  volume = {104},
  issue = {15},
  pages = {155428},
  numpages = {16},
  year = {2021},
  month = {Oct},
  publisher = {American Physical Society},
  doi = {10.1103/PhysRevB.104.155428},
  url = {https://link.aps.org/doi/10.1103/PhysRevB.104.155428}
}

@article{salimian2021gate,
  author={Salimian, Sedighe and Carrega, Matteo and Verma, Isha and Zannier, Valentina and Nowak, Micha{\l} P and Beltram, Fabio and Sorba, Lucia and Heun, Stefan},
  title={Gate-controlled supercurrent in ballistic InSb nanoflag Josephson junctions},
  journal={Applied Physics Letters},
  volume={119},
  number={21},
  year={2021},
  publisher={AIP Publishing},
url={https://pubs.aip.org/aip/apl/article/119/21/214004/40738/Gate-controlled-supercurrent-in-ballistic-InSb}
}

@article{PhysRevB.95.174515,
  author = {Setiawan, F. and Cole, William S. and Sau, Jay D. and Das Sarma, S.},
  title = {Transport in superconductor--normal metal--superconductor tunneling structures: Spinful $p$-wave and spin-orbit-coupled topological wires},
  journal = {Phys. Rev. B},
  volume = {95},
  issue = {17},
  pages = {174515},
  numpages = {16},
  year = {2017},
  month = {May},
  publisher = {American Physical Society},
  doi = {10.1103/PhysRevB.95.174515},
  url = {https://link.aps.org/doi/10.1103/PhysRevB.95.174515}
}

@article{PhysRevResearch.4.043087,
  author = {Setiawan, F. and Hofmann, Johannes},
  title = {Analytic approach to transport in superconducting junctions with arbitrary carrier density},
  journal = {Phys. Rev. Res.},
  volume = {4},
  issue = {4},
  pages = {043087},
  numpages = {24},
  year = {2022},
  month = {Nov},
  publisher = {American Physical Society},
  doi = {10.1103/PhysRevResearch.4.043087},
  url = {https://link.aps.org/doi/10.1103/PhysRevResearch.4.043087}
}

@article{PhysRevB.107.245304,
  author = {Banerjee, A. and Lesser, O. and Rahman, M. A. and Wang, H.-R. and Li, M.-R. and Kringh\o{}j, A. and Whiticar, A. M. and Drachmann, A. C. C. and Thomas, C. and Wang, T. and Manfra, M. J. and Berg, E. and Oreg, Y. and Stern, Ady and Marcus, C. M.},
  title = {Signatures of a topological phase transition in a planar Josephson junction},
  journal = {Phys. Rev. B},
  volume = {107},
  issue = {24},
  pages = {245304},
  numpages = {12},
  year = {2023},
  month = {Jun},
  publisher = {American Physical Society},
  doi = {10.1103/PhysRevB.107.245304},
  url = {https://link.aps.org/doi/10.1103/PhysRevB.107.245304}
}

@article{Pekerten2022:PRB,
  author = {Pekerten, Baris and Pakizer, Joseph D. and Hawn, Benjamin and Matos-Abiague, Alex},
  title = {Anisotropic topological superconductivity in Josephson junctions},
  journal = {Phys. Rev. B},
  volume = {105},
  issue = {5},
  pages = {054504},
  numpages = {13},
  year = {2022},
  month = {Feb},
  publisher = {American Physical Society},
  doi = {10.1103/PhysRevB.105.054504},
  url = {https://link.aps.org/doi/10.1103/PhysRevB.105.054504}
}

@article{PhysRevB.107.035435,
  author = {Luethi, Melina and Laubscher, Katharina and Bosco, Stefano and Loss, Daniel and Klinovaja, Jelena},
  title = {Planar Josephson junctions in germanium: Effect of cubic spin-orbit interaction},
  journal = {Phys. Rev. B},
  volume = {107},
  issue = {3},
  pages = {035435},
  numpages = {19},
  year = {2023},
  month = {Jan},
  publisher = {American Physical Society},
  doi = {10.1103/PhysRevB.107.035435},
  url = {https://link.aps.org/doi/10.1103/PhysRevB.107.035435}
}

@article{bychkov1984oscillatory,
  author={Bychkov, Yu A and Rashba, Emmanuel I},
  title={Oscillatory effects and the magnetic susceptibility of carriers in inversion layers},
  journal={Journal of physics C: Solid state physics},
  volume={17},
  number={33},
  pages={6039},
  year={1984},
  publisher={IOP Publishing},
url={https://iopscience.iop.org/article/10.1088/0022-3719/17/33/015}
}

@article{dresselhaus1955spin,
  author={Dresselhaus, Gene},
  title={Spin-orbit coupling effects in zinc blende structures},
  journal={Physical Review},
  volume={100},
  number={2},
  pages={580},
  year={1955},
  publisher={APS},
url={https://journals.aps.org/pr/abstract/10.1103/PhysRev.100.580}
}

@article{PhysRevB.56.892,
  author = {Tanaka, Yukio and Kashiwaya, Satoshi},
  title = {Theory of Josephson effects in anisotropic superconductors},
  journal = {Phys. Rev. B},
  volume = {56},
  issue = {2},
  pages = {892--912},
  numpages = {0},
  year = {1997},
  month = {Jul},
  publisher = {American Physical Society},
  doi = {10.1103/PhysRevB.56.892},
  url = {https://link.aps.org/doi/10.1103/PhysRevB.56.892}
}

@article{PhysRevLett.105.097002,
  author = {Tanaka, Yukio and Mizuno, Yoshihiro and Yokoyama, Takehito and Yada, Keiji and Sato, Masatoshi},
  title = {Anomalous Andreev Bound State in Noncentrosymmetric Superconductors},
  journal = {Phys. Rev. Lett.},
  volume = {105},
  issue = {9},
  pages = {097002},
  numpages = {4},
  year = {2010},
  month = {Aug},
  publisher = {American Physical Society},
  doi = {10.1103/PhysRevLett.105.097002},
  url = {https://link.aps.org/doi/10.1103/PhysRevLett.105.097002}
}

@article{PhysRevB.97.205134,
  author = {Zhu, Xiaoyu},
  title = {Tunable Majorana corner states in a two-dimensional second-order topological superconductor induced by magnetic fields},
  journal = {Phys. Rev. B},
  volume = {97},
  issue = {20},
  pages = {205134},
  numpages = {7},
  year = {2018},
  month = {May},
  publisher = {American Physical Society},
  doi = {10.1103/PhysRevB.97.205134},
  url = {https://link.aps.org/doi/10.1103/PhysRevB.97.205134}
}

@article{PhysRevB.109.115413,
  author = {Liu, Lizhou and Miao, Chengming and Tang, Hanzhao and Zhang, Ying-Tao and Qiao, Zhenhua},
  title = {Magnetically controlled topological braiding with Majorana corner states in second-order topological superconductors},
  journal = {Phys. Rev. B},
  volume = {109},
  issue = {11},
  pages = {115413},
  numpages = {8},
  year = {2024},
  month = {Mar},
  publisher = {American Physical Society},
  doi = {10.1103/PhysRevB.109.115413},
  url = {https://link.aps.org/doi/10.1103/PhysRevB.109.115413}
}

@article{PhysRevB.109.205158,
  author = {Winblad, Aidan and Chen, Hua},
  title = {Superconducting triangular islands as a platform for manipulating Majorana zero modes},
  journal = {Phys. Rev. B},
  volume = {109},
  issue = {20},
  pages = {205158},
  numpages = {7},
  year = {2024},
  month = {May},
  publisher = {American Physical Society},
  doi = {10.1103/PhysRevB.109.205158},
  url = {https://link.aps.org/doi/10.1103/PhysRevB.109.205158}
}

@article{PhysRevB.110.L041110,
  author = {Nagae, Yutaro and Schnyder, Andreas P. and Tanaka, Yukio and Asano, Yasuhiro and Ikegaya, Satoshi},
  title = {Multilocational Majorana zero modes},
  journal = {Phys. Rev. B},
  volume = {110},
  issue = {4},
  pages = {L041110},
  numpages = {7},
  year = {2024},
  month = {Jul},
  publisher = {American Physical Society},
  doi = {10.1103/PhysRevB.110.L041110},
  url = {https://link.aps.org/doi/10.1103/PhysRevB.110.L041110}
}

@article{Pekerten2024:APL,
    author = {Pekerten, Barış and Brandão, David S. and Bussiere, Bailey and Monroe, David and Zhou, Tong and Han, Jong E. and Shabani, Javad and Matos-Abiague, Alex and Žutić, Igor},
    title = "{Beyond the standard model of topological Josephson junctions: From crystalline anisotropy to finite-size and diode effects}",
    journal = {Appl. Phys. Lett.},
    volume = {124},
    number = {25},
    pages = {252602},
    year = {2024},
    month = {06},
    doi = {10.1063/5.0214920},
    url = {https://doi.org/10.1063/5.0214920}
}

@article{Marra2024:JAP,
    author = {Marra, Pasquale},
    title = "{Majorana nanowires for topological quantum computation}",
    journal = {Journal of Applied Physics},
    volume = {132},
    number = {23},
    pages = {231101},
    year = {2022},
    month = {12},
    issn = {0021-8979},
    doi = {10.1063/5.0102999},
    url = {https://doi.org/10.1063/5.0102999},
}

@article{Kovalev2022:JAP,
    author = {Güngördü, Utkan and Kovalev, Alexey A.},
    title = "{Majorana bound states with chiral magnetic textures}",
    journal = {Journal of Applied Physics},
    volume = {132},
    number = {4},
    pages = {041101},
    year = {2022},
    month = {07},
    issn = {0021-8979},
    doi = {10.1063/5.0097008},
    url = {https://doi.org/10.1063/5.0097008},
}

@article{
Lian2018:PNAS,
    author = {Biao Lian  and Xiao-Qi Sun  and Abolhassan Vaezi  and Xiao-Liang Qi  and Shou-Cheng Zhang },
    title = {Topological quantum computation based on chiral Majorana fermions},
    journal = {Proceedings of the National Academy of Sciences},
    volume = {115},
    number = {43},
    pages = {10938-10942},
    year = {2018},
    doi = {10.1073/pnas.1810003115},
    URL = {https://www.pnas.org/doi/abs/10.1073/pnas.1810003115}
}

@article{menard2019:Nature,
  author={M{\'e}nard, Gerbold C and Mesaros, Andrej and Brun, Christophe and Debontridder, Fran{\c{c}}ois and Roditchev, Dimitri and Simon, Pascal and Cren, Tristan},
  title={Isolated pairs of Majorana zero modes in a disordered superconducting lead monolayer},
  journal={Nature communications},
  volume={10},
  number={1},
  pages={2587},
  year={2019},
  publisher={Nature Publishing Group UK London},
  url={https://www.nature.com/articles/s41467-019-10397-5#citeas}
}

@article{Beenakker2020:SP,
  author={C. W. J. Beenakker and Dmytro Oriekhov},
  title={Shot noise distinguishes Majorana fermions from vortices injected in the edge mode of a chiral p-wave superconductor},
  journal={SciPost Physics},
  year={2020},
  url={https://api.semanticscholar.org/CorpusID:221995636}
}

@article{wu2017:PRB,
  author={Wu, Chien-Te and Anderson, Brandon M and Hsiao, Wei-Han and Levin, K},
  title={Majorana zero modes in spintronics devices},
  journal={Physical Review B},
  volume={95},
  number={1},
  pages={014519},
  year={2017},
  publisher={APS},
url={https://journals.aps.org/prb/abstract/10.1103/PhysRevB.95.014519}
}

@article{Daido2017:PRB,
  author = {Daido, Akito and Yanase, Youichi},
  title = {Majorana flat bands, chiral Majorana edge states, and unidirectional Majorana edge states in noncentrosymmetric superconductors},
  journal = {Phys. Rev. B},
  volume = {95},
  issue = {13},
  pages = {134507},
  numpages = {11},
  year = {2017},
  month = {Apr},
  publisher = {American Physical Society},
  doi = {10.1103/PhysRevB.95.134507},
  url = {https://link.aps.org/doi/10.1103/PhysRevB.95.134507}
}

@article{Sau2010:PRL,
  author = {Sau, Jay D. and Lutchyn, Roman M. and Tewari, Sumanta and Das Sarma, S.},
  title = {Generic New Platform for Topological Quantum Computation Using Semiconductor Heterostructures},
  journal = {Phys. Rev. Lett.},
  volume = {104},
  issue = {4},
  pages = {040502},
  numpages = {4},
  year = {2010},
  month = {Jan},
  publisher = {American Physical Society},
  doi = {10.1103/PhysRevLett.104.040502},
  url = {https://link.aps.org/doi/10.1103/PhysRevLett.104.040502}
}

@article{Pekerten2024:PRB,
  author = {Pekerten, Bar{\i}{\c{s}} and Brand\~ao, David and Elfeky, Bassel Heiba and Zhou, Tong and Han, Jong E. and Shabani, Javad and Žutić, Igor},
  title = {Microwave signatures of topological superconductivity in planar Josephson junctions},
  journal = {Phys. Rev. B},
  volume = {110},
  issue = {6},
  pages = {L060513},
  numpages = {8},
  year = {2024},
  month = {Aug},
  publisher = {American Physical Society},
  doi = {10.1103/PhysRevB.110.L060513},
  url = {https://link.aps.org/doi/10.1103/PhysRevB.110.L060513}
}

@article{Paudel2205:PReApplied,
  title = {Disorder effects in planar semiconductor-superconductor structures: Majorana wires versus Josephson junctions},
  author = {Paudel, Purna P. and Smith, Nathan O. and Stanescu, Tudor D.},
  journal = {Phys. Rev. Appl.},
  volume = {23},
  issue = {3},
  pages = {034068},
  numpages = {27},
  year = {2025},
  month = {Mar},
  publisher = {American Physical Society},
  doi = {10.1103/PhysRevApplied.23.034068},
  url = {https://link.aps.org/doi/10.1103/PhysRevApplied.23.034068}
}

@misc{sharma2024:arxiv,
      title={Multiple Majorana bound states and their resilience against disorder in planar Josephson junctions}, 
      author={Pankaj Sharma and Narayan Mohanta},
      year={2024},
      eprint={2409.07532},
      archivePrefix={arXiv},
      primaryClass={cond-mat.supr-con},
      url={https://arxiv.org/abs/2409.07532}, 
}

@article{Haim2019:PRL,
  title = {Benefits of Weak Disorder in One-Dimensional Topological Superconductors},
  author = {Haim, Arbel and Stern, Ady},
  journal = {Phys. Rev. Lett.},
  volume = {122},
  issue = {12},
  pages = {126801},
  numpages = {5},
  year = {2019},
  month = {Mar},
  publisher = {American Physical Society},
  doi = {10.1103/PhysRevLett.122.126801},
  url = {https://link.aps.org/doi/10.1103/PhysRevLett.122.126801}
}

@article{Laubscher2025:PRB,
  title = {Detection of Majorana zero modes bound to Josephson vortices in planar superconductor--topological insulator--superconductor junctions},
  author = {Laubscher, Katharina and Sau, Jay D.},
  journal = {Phys. Rev. B},
  volume = {111},
  issue = {23},
  pages = {235442},
  numpages = {13},
  year = {2025},
  month = {Jun},
  publisher = {American Physical Society},
  doi = {10.1103/PhysRevB.111.235442},
  url = {https://link.aps.org/doi/10.1103/PhysRevB.111.235442}
}

@article{Escribano2025:PRB,
  title = {Phase-controlled minimal Kitaev chain in multiterminal Josephson junctions},
  author = {Escribano, Samuel D. and Dahl, Anders Enevold and Flensberg, Karsten and Oreg, Yuval},
  journal = {Phys. Rev. B},
  volume = {111},
  issue = {19},
  pages = {195433},
  numpages = {16},
  year = {2025},
  month = {May},
  publisher = {American Physical Society},
  doi = {10.1103/PhysRevB.111.195433},
  url = {https://link.aps.org/doi/10.1103/PhysRevB.111.195433}
}

@article{Flensberg2011:PRL,
  title = {Non-Abelian Operations on Majorana Fermions via Single-Charge Control},
  author = {Flensberg, Karsten},
  journal = {Phys. Rev. Lett.},
  volume = {106},
  issue = {9},
  pages = {090503},
  numpages = {4},
  year = {2011},
  month = {Mar},
  publisher = {American Physical Society},
  doi = {10.1103/PhysRevLett.106.090503},
  url = {https://link.aps.org/doi/10.1103/PhysRevLett.106.090503}
}

@article{vanHeck2012:NJP,
doi = {10.1088/1367-2630/14/3/035019},
url = {https://doi.org/10.1088/1367-2630/14/3/035019},
year = {2012},
month = {mar},
publisher = {IOP Publishing},
volume = {14},
number = {3},
pages = {035019},
author = {van Heck, B and Akhmerov, A R and Hassler, F and Burrello, M and Beenakker, C W J},
title = {Coulomb-assisted braiding of Majorana fermions in a Josephson junction array},
journal = {New J. Phys.},
}

\end{document}